\documentclass[12pt,preprint]{aastex}
\begin{document}
\title{A Far Ultraviolet Study of the Nova-like V794 Aquilae  
\altaffilmark{1}}

\author{Patrick Godon\altaffilmark{2}, Edward M. Sion}
\affil{Dept. of Astronomy \& Astrophysics,
Villanova University,
Villanova, PA 19085,
patrick.godon@villanova.edu, edward.sion@villanova.edu}

\author{Paul Barrett}
\affil{United States Naval Observatory,  
Washington, DC 20392, 
barrett.paul@usno.navy.mil}

\author{Paula Szkody}
\affil{Department of Astronomy,
University of Washington,
Seattle, WA 98195,
szkody@astro.washington.edu}

\altaffiltext{1}
{Based on observations made with the NASA-CNES-CSA Far Ultraviolet 
Spectroscopic
Explorer. {\it{FUSE}} is operated for NASA by the Johns 
Hopkins University under NASA contract NAS5-32985} 
\altaffiltext{2}
{Visiting at the Space Telescope Science Institute, Baltimore, MD 21218,
USA, godon@stsci.edu}

\begin{abstract}

V794 Aql was observed with 
the Hubble Space Telescope/Space Telescope Imaging Spectrograph
({\it{HST}}/STIS) on August 28, 2003, and with  
the Far Ultraviolet Spectroscopic
Explorer ({\it{FUSE}}) on May 13, 2004. 
In both observations V794 Aql was found in a relatively high state
with the same flux level. 
The {\it{FUSE}} spectrum exhibits heavy interstellar 
hydrogen absorption features that implies
$N(H_I)=4.5 \times 10^{20}$cm$^{-2}$, $N(H_2)=3 \times 10^{17}$cm$^{-2}$, 
and a reddening value E(B-V)=0.08. Inspection of the 
existing archival {\it{IUE}} spectra also indicates that the reddening
could be large $\sim 0.1-0.2$. 
We present here a spectral analysis of 
the dereddened {\it{FUSE}} and {\it{HST}}/STIS spectra separately 
and combined together assuming E(B-V)=0.1 \& 0.2.  
Overall, we find that the model fits are in much better agreement
with the dereddened spectra when E(B-V) is large, as excess emission in
the longer wavelengths render the slope of the observed spectra almost
impossible to fit, unless E(B-V)=0.2 .  

For a $M_{wd} \approx 0.9 M_{\odot}$ 
the best fit is an accretion disk with a mass accretion rate 
$\dot{M}= 10^{-8.5}-10^{-8.0} M_{\odot}$ yr$^{-1}$ with
an inclination $i=60\deg$ when assuming E(B-V)=0.2, and 
$\dot{M}$ decreases by one order of magnitude when assuming
E(B-V)=0.1. The distance to the system is $d\approx 690 \pm 105$pc
for the E(B-V)=0.2 case and it decreases to $d \approx 310 - 834$pc
for E(B-V)=0.1. 
The best fit accretion disk model is obtained for E(B-V)=0.2 . 
A single white dwarf model leads to a rather 
hot temperature ($30,000K < T_{wd} < 55,000K$ depending on the 
assumptions) but does not provide a fit as good as the accretion disk model. 
A combination of a white dwarf plus a disk does not lead to a better
fit. The same best fit disk model is consistently obtained when fitting
the {\it{FUSE}} and {\it{HST}}/STIS spectra 
individually and when combined together,
implying therefore that the disk model is the best fit not only in the
least $\chi^2$ sense, but also as a consistent 
solution across a large wavelength
span of observation. This is not the case with the single white model fitting 
which leads to a different (and therefore inconsistent) 
temperature for each different spectrum ({\it{FUSE}}, 
STIS and {\it{FUSE}}+STIS).

\end{abstract}

Subject headings: accretion, accretion disks - novae, cataclysmic variables 
- stars: individual (V794 Aquilae) - ultraviolet: stars - white dwarfs  

\section{Introduction}

Cataclysmic variables (CVs) are short-period, semi-detached binary systems
consisting of an accreting white dwarf (WD) star (the primary) and a low-mass
main-sequence star (the secondary) as the Roche lobe-filling mass donor
\citep{war95}. In non magnetic CV systems, the mass is accreted by means
of an accretion disk reaching all the way down to the surface of the WD. 
On-going accretion at a low rate (quiescence) is interrupted every few weeks
to months by intense accretion (outburst) of days to weeks - a dwarf nova (DN)
accretion event. CV systems are divided in sub-classes according
to the durations, occurrence and amplitude of 
their outburst: e.g. dwarf nova systems
(DNs) spend most of their time in the quiescent state, 
while nova-like systems (NLs)
are found mostly in the high outburst state. Both DN and non-magnetic NL  
systems exhibit emission from the accretion disk during the high state.  

Far ultraviolet (FUV) observations have shown 
that some WDs in CVs can be directly
viewed in the UV when the accretion disk is not dominant
(as ealy as \citet{mat84}). Consequently, much effort has
gone to observe the systems with low mass transfer rates. In those
systems the WDs are hotter than single field WDs and their temperature
increases with orbital period, as higher accretion rates are found
in longer period systems \citet{sio99}. 
Interestingly enough, while the binary period
of the CV systems ranges between a fraction of an hour (e.g. AM CVn
systems) up to $\approx 2$ day (e.g. GK Per), there is a gap between
about 2 and 3h where almost no system is found (hereafter the ``period
gap''). However, at a given orbital period the
accretion rate can vary by a large amount and unfortunately
there are not many data points for long period systems above the
period gap ($>$3hrs). In the accretion disk limit cycle \citep{can93}, 
material accumulates in the accretion disk 
during quiescence and accretes onto the WD during outburst. In order
for an outburst to occur, the accretion rate must be below a critical
value for a given orbital period. 
DNe at periods above the gap should have accretion rates below 
the limit, the rates for Z Cam are very close
to the critical value and the novalikes have rates above the limit. 
Thus, above the period gap one finds
3 classes  of systems : U Gem (DN) type that undergo outbursts;
Z Cam (DN) type which have standstills where they remain at about
one magnitude below their outburst level for long times; 
and nova-like systems that are in a permanent high state. 

Among these high accretion NL systems, only the underlying 
WDs of DW UMa \citep{kni00}, TT Ari \citep{gan99}  and MV Lyr \citep{hoa04}
have been studied for temperature, gravity, rotation and chemical 
abundances information.  
All the other NLs that have been observed have
been caught in a high state. The spectra of NLs (in permanent high state) 
consist of H emission lines which may of may not
be superimposed on broad shallow absorption features;
and the brightness fluctuates about some mean value, deviating up and
down irregularly by no more than about 1 magnitude. The spectra in the near 
UV and optical is consistent with that of an accretion disk in outburst. 
While NLs are characterized by an approximately steady, high rate of
mass transfer (and consequently a high luminous accretion disk - a
permanent high state), there is a class of NLs  
(namely, the VY Sculptoris systems) in which the prolonged state
of high accretion is interrupted unpredictably by low brightness
states of little or no accretion when the disk greatly shrinks or vanishes and 
the underlying accretion heated hot WD is exposed, as it is the case 
for MV Lyr \citep{hoa04}. 
The VY Scl systems are apparently all disk systems
with negligible (magnetic) accretion at the poles. 
V794 Aql is a nova-like system belonging to the VY Sculptoris class,
or at least it appears to have observational properties
consistent with this class. 
The VY Scl stars fall above the period gap, all in the range 3-4hr, 
except for V751 Cyg with $P_{orb} > 4$h. 

\subsection{V794 Aql}

The VY Scl have all extreme
(3-magnitude) brightness variations in the high state
and unexpectedly fall to an extreme low state (more than 5 magnitudes). 
V794 Aql has been seen to vary erratically
between photographic magnitude 14 and 17 on time scales of days to
years \citep{szk81},  
and \citet{hon85} observed V794 Aql in an 
unprecedented low state with magnitude 20! Photometrically speaking,
V794 Aql seems to be 
one of the most active cataclysmic variables with brightness
variations of up to 0.5 mag occurring on a timescale of minutes
(and smaller flares on timescales of tens of seconds, \citep{war82}).  
The brightness variations in VY Scl systems 
appear to be random in amplitude, shape and
recurrence interval. However, V794 Aql is an exception:  
photometric monitoring of V794 Aql has revealed an unusual 
type of light curve \citep{hon94,hon98}, in which the uniform 
decline from high to low state is interupted by an abrupt 
return to the high state, giving the light curve a distinctive 
'sawtooth' appearance. This is a type of photometric variation in 
CVs which has never been reported. 

V794 Aql was spectroscopically observed in high and low state in the optical 
\citep{szk81,hon85,hon98}, in the UV with {\it{IUE}}
\citep{szk88,lad91} and in the X-ray with Einstein \citep{szk88}, 
and in respects other than the character of its long-term light
curve, V794 Aql was found to have a photometric and spectroscopic behavior 
very similar to other members of the NL VY Sculptoris class,
such as MV Lyrae and TT Arietis. 

In the X-ray, V794 Aql was observed  
near a high state in terms of its optical range.
The system had a flux of $4.5 \times 10^{-12}$ ergs$~$cm$^{-2}$s$^{-1}$
(in the 0.1-4.5 keV band), corresponding to an X-ray luminosity of
$L_X=5.4 \times 10^{30}$ergs$~$s$^{-1}$, for a distance d=100pc.  
The source was detected in both the soft (0.1-0.5 keV) and hard
(0.5-4.5 keV) energy bands, and the hardness ratio was found to be
$11 \pm 2$ (though the apparent hardness can be affected by the absorption 
of soft X-rays either at the source itself or in the ISM; and ISM 
absorption might not be negligible for V794 Aql - see section 2.2).  
Simultaneous ground-based optical observations showed blue colors,
strong H emission with a flat Balmer decrement, plus strong
He {\small{I}} and He {\small{II}} $\lambda$4686 emission.  
These {\it{IUE}}, X-ray, as well as optical spectra show relatively strong
emission lines of intermediate excitation,
similar to those seen in other VY Sculptoris stars in both the low state 
and the high state. 
In the low state reported by \citet{hon85} the accretion disk
emission lines were replaced by very narrow Balmer emission,
in a behavior similar to (but more extreme than) that of MV Lyr 
(e.g. \citet{hoa04}) and a few other NL systems in which the 
accretion apparently switches off for brief periods of time. 

Though V794 Aql has many of the properties of a typical disk VY Scl
star, there has been no direct evidence yet that a disk exists
in this system. V794 Aql cannot be classified as a Z Cam system,
as the sawtooths are consistently much fainter than the
occasional intervals of steady brightness characteristic of Z Cam
systems. If V794 Aql is a Z Cam system, then it is certainly 
a very unusual one.   

\citet{szk88} studied V794 Aql from {\it{IUE}}  
spectra, however the low resolution and geocoronal contamination
of the {\it{IUE}} spectra prevented accurate determination of the
temperature and the low response shortward of Ly$\alpha$ prevented
any observation of a rising continuum from the hot WD.  
This study estimated
a white dwarf temperature near 50,000K, while the accretion 
rate has been estimated
to be around $10^{-8}M_{\odot}$yr$^{-1}$ \citep{hon94} - 
based on the long term optical variations of the
system. This NL system has the highest mass accretion rate, 
above the limit for outbursts. Because of its inclination, 
both the hot WD and the inner disk might be observable. 

More recently an {\it{HST}}/STIS snapshot of V794 Aql was obtained, 
and the system was also observed in the FUV with {\it{FUSE}} under a
Cycle 4 program.  
In this paper, we report a spectral analysis of
the {\it{FUSE}} and {\it{HST}}/STIS spectra of V794 Aql 
using accretion disk models, photosphere
models, and models combining white dwarfs and accretion disks.  
Our objectives are
to identify the source(s) of the FUV radiation, derive
the properties of the WD (if possible), the accretion disk, and
characterize the hot component in the system.

\section{The Observations}

We report here all the ultraviolet spectra of V794 Aql. 
This includes 5 {\it{IUE}} archival
spectra, one {\it{HST}}/STIS spectrum and one 
{\it{FUSE}} spectrum. While we model mainly the {\it{FUSE}}
and {\it{HST}}/STIS spectra, we also use the 
{\it{IUE}} spectra for flux comparison to estimate
the relative state in which the system was found during each observation. 
For all the spectra we identify 
the emission and absorption features with a 
particular emphasis on the estimate of the
molecular and atomic hydrogen column densities using the {\it{FUSE}} 
and {\it{HST}}/STIS (which covers the Ly $\alpha$ region).  
We use the {\it{IUE}} spectra to carry out an assessment 
of the reddening of the system. 
The reason for questioning  the zero color excess value of \citet{ver87}  
is in part the {\it{a posteriori}} fact that 
the slope of the continuum (both in the {\it{FUSE}} 
and STIS spectra) cannot be  easily  
matched with the model fits when assuming a small E(B-V) value. 
We obtain a large residual in the
longer wavelength indicating a net excess of flux in the 'red' part 
of the spectrum. 
This discrepancy basically disappears when we deredden the spectra assuming
$E(B-V)=0.2 $. In addition, the hydrogen column density we 
obtain from the {\it{FUSE}} and STIS spectra leads to  
a reddening of E(B-V)=0.08 using the relation of \citet{boh78}, 

\subsection{The {\it{IUE}} Archival Spectra}

The {\it{IUE}} archive contains three good observations 
of V794 Aql with the
short-wavelength prime (SWP) camera through the large aperture at low
dispersion (see Table 2 and Figure 1). 
One additional {\it{IUE}} SWP spectrum (SWP 14708)
has an extremely low flux
level and  a very low  signal to noise (S/N) ratio. 
This spectrum has been disregarded here as it has basically
no continuum and cannot be used. 
The {\it{IUE}} SWP spectra have a wavelength binning of 1.68 \AA , covering the 
wavelength range 1150 to 1980\AA. 
There is also a data set obtained with the long-wavelength 
camera LWR 11782, which we use to assess the reddening of the system. 
The {\it{IUE}} LWR spectrum has a wavelength binning of 1.67 \AA , 
and covers the wavelength range 1850-3350\AA . 
All the {\it{IUE}} NEWSIPS spectra were flux-calibration corrected using the 
Massa-Fitzpatrick corrections \citep{mas00}. 
These spectra were analyzed in detail by \citet{szk88,lad91}. 

The SWP 28501 spectrum was obtained when the system had a visual 
magnitude of 17-18, a relatively low state, 
though not the lowest state of V794 Aql 
(magnitude 20 reported by \citet{hon85}).  
The spectrum SWP 15266 has a flux only slightly larger than SWP 28501, but 
in addition it exhibits some broad emission lines. The most
prominent ones are 
$N_V$ (1238/1243\AA),
$Si_{IV}$ (1394/1403\AA),  
$C_{IV}$ (1548/1550\AA), 
and 
$He_{II}$ (1640\AA\ Balmer $\alpha$). 
With its broad emission lines characteristic of nova-like systems,
the spectrum of V794 Aql show evidence of hot gas.
The SWP 50754 shows the system in a relatively higher state with 
emission and absorption features, however, the S/N of this spectra is very low. 

The presently accepted value for the reddening of V794 Aql 
is $E(B-V)=0$ (e.g. \citet{lad91,bru94}, based on the work of \citet{ver87}).  
However, the heavy interstellar hydrogen absorption in its {\it{FUSE}} spectrum
(see next subsection) 
implies that the source is either located pretty far away, or masked by an
ISM cloud, or some cold circumstellar matter, and consequently 
it is probably also affected
by dust, therefore making the assumption $E(B-V)=0$ very unlikely. 
In his pioneer work,  \citet{ver87} carried out an analysis   
of the broad (several hundreds \AA\ ) absorption feature around 2175 \AA\ 
of the {\it{IUE}} spectra of 51 CV systems to determine their reddening value
E(B-V). The reddening of the 51 systems was assessed by dereddening
each spectrum assuming increasing values of E(B-V) (e.g. 0.05, 0.10, 0.15..)
until the broad absorption dip (between $\approx $2000\AA\ and 
$\approx $2400\AA ) disappeared 
completely to the ``eye'' (i.e. the acceptable range of values of E(B-V)  
were found by visual inspection).   

Following \citet{ver87} we first combine the {\it{IUE}} spectra SWP 28501 and 
LWR 11782 binned at 20\AA\ (we have  chosen the SWP 28501 spectrum 
as it matches the flux level and shape of LWR 11782). 
The spectra are very noisy, including a 'hot pixel' at 2200\AA . 
This combined spectrum is then dereddened assuming values 
E(B-V)=0.10, 0.20 and 0.30.  
%
In Figure 2 we plot the AB-magnitude \citep{oke74} for each 
dereddened spectrum for visual inspection.  

The broad absorption feature is more apparent in the shortest wavelengths 
between 2000\AA\ and 2200\AA\ in the E(B-V)=0 spectrum, an indication that
the reddening is not zero. 
Obviously for E(B-V)=0.30 there is excess emission around $2200 \pm 300$\AA . 
For E(B-V)=0.10 and 0.20 the absorption feature is not that apparent as the
flux seems to level off,   
an indication that the reddening is probably around $\approx 0.10$
but could be as large as 0.20. Since we have only one {\it{IUE}} LWR spectrum
of V794 Aql, and it is extremely noisy, it is difficult to say more about
the reddening. In addition,
we see how difficult
it can be to assess 'how much' of the broad absorption feature is present
when it is left only to a visual inspection to decide, as one cannot
differentiate between noise and signal. This is probably
the reason why \citet{ver87} assumed an error of 0.1 for V794 Aql.
The difference between our reddening assessment and Verbunt's might
reside in the fact that Verbunt co-added all the SWP spectra together,
even though only the SWP 28501 spectrum matches the LWR 11782 spectrum.  
Also, since the estimate is carried out by visual inspection, the
subjectivity of the observer is also a decisive factor. 

\subsection{The {\it{FUSE}} Observations}

{\it{FUSE}} is a low-earth orbit satellite, launched in June 1999. Its optical
system consists of four optical telescopes (mirrors), each separately
connected to a different Rowland spectrograph. The four diffraction
gratings of the four Rowland spectrographs produce four independent 
spectra on two microchannel plates. Two mirrors and two 
gratings are coated with SiC to provide wavelength coverage below 
1020 \AA, while the other
two mirrors and gratings are coated with Al and LiF overcoat.  The
Al+LiF coating provides about twice the reflectivity of SiC at
wavelengths $>$1050 \AA, and very little reflectivity below 1020 \AA\
(hereafter the SiC1, SiC2, LiF1 and LiF2 channels).  

A TIME TAG {\it{FUSE}} spectrum 
(D1440101) of V794 Aql was obtained
starting on 2003, May 13, (at 19:50:22) 
with a total observing duration covering 6 individual spacecraft 
orbits through the LWRS. 
The system was in a relatively high state at the time
of the observation.  The data were processed
with CalFUSE version 3.0 totaling 13,066s of good exposure time.
The main change from previous versions of CalFUSE is that
now the data are maintained as a photon list (the intermediate
data file - IDF) throughout the pipeline. Bad photons are flagged 
but not discarded, so the user can examine, filter, and combine
data without re-running the pipeline. A number of design changes
enable the new pipeline to run faster and use less disk space than 
before. Processing time with CalFUSE has decreased by a factor of up to 10. 
In this version, event bursts are automatically taken care of. 
Event bursts are short periods during an exposure when high count 
rates are registered on one of more detectors. The bursts exhibit  
a complex pattern on the detector, their cause, however, is yet unknown 
(it has been confirmed that they are not detector effects). 

During the observations, Fine Error Sensor A, which images the LiF 1 aperture
was used to guide the telescope. The spectral regions covered by the
spectral channels overlap, and these overlap regions are then used to
renormalize the spectra in the SiC1, LiF2, and SiC2 channels to the flux in
the LiF1 channel. We then produced a final spectrum that covers almost the
full {\it{FUSE}} wavelength range $905-1182$ \AA. 
The low sensitivity portions of each channel were discarded.
%
%
Here we took particular care to discard the portion of the spectrum 
where the so-called {\it{worm}} 'crawls', 
which deteriorates LiF1 longward of $1125$ \AA\  
(CalFUSE cannot correct target fluxes for this effect).  
Because of this the $1182 - 1187$ \AA\ region was lost
(see the {\it{FUSE}} Instrument and Data Handbook or
section 2.1 in \citet{god06}).  
We combined the individual exposures and channels to create a
time-averaged spectrum with a linear, $0.1$ \AA\ dispersion, weighting
the flux in each output datum by the exposure time and sensitivity of the
input exposure and channel of origin. 

V794 Aql, with a flux of a few $ 10^{-14}$ergs$~$s$^{-1}$cm$^{-2}$\AA$^{-1}$,
is actually a relatively weak source. 
The procedure we used to process the {\it{FUSE}} data of V794 Aql is the same  
as the one we used in our previous {\it{FUSE}} analysis
of CVs, such as e.g. RU Peg \&  SS Aur \citep{sio04}, VW Hyi \citep{god04}, 
or WW Cet \citep{god06}.  \\ 

{\bf{Interstellar Absorption Toward V794 Aql}} 

A single look at the {\it{FUSE}} spectrum 
of V794 Aql (Figure 3) immediately reveals
that the continuum is moderately affected by hydrogen absorption:
the continuum is basically sliced (at almost equal intervals) by interstellar
hydrogen lines starting at wavelengths around 1110\AA\ and continuing 
towards shorter wavelengths all the way down to the hydrogen cut-off around 
915\AA\ . In Table 3 we identified all the absorption and emission lines
of metals in the {\it{FUSE}} spectrum of V794 Aql. In Table 4 we identified the
most prominent molecular hydrogen absorption lines by their band 
(Werner or Lyman), upper vibrational level (1-16), and rotational transition
(R, P, or Q) with lower rotational state (J=1,2,3).

Next, we model the ISM hydrogen absorption lines to assess   
the atomic and molecular column densities. 
We use a custom spectral fitting package to estimate the temperature
and density of the interstellar absorption lines of atomic and
molecular hydrogen.  The ISM model assumes that the temperature, bulk
velocity, and turbulent velocity of the medium are the same for all
atomic and molecular species, whereas the densities of atomic and
molecular hydrogen, and the ratios of deuterium to hydrogen and metals
(including helium) to hydrogen can be adjusted independently. The
model uses atomic data of \citet{mor00,mor03} and molecular data of
\citet{abg00}. The optical depth calculations of molecular 
hydrogen have been checked against those of \citet{mcc03}.

For V794 Aql, the ratios of metals to hydrogen and deuterium to
hydrogen are fixed at 0 and $2 \times 10^{-5}$, respectively, 
because of the low
signal-to-noise ratio data.  The wings of the atomic lines are used to
estimate the density of atomic hydrogen and the depth of the
unsaturated molecular lines for molecular hydrogen.  The temperature
and turbulent velocity of the medium are primarily determined from the
lines of molecular hydrogen when the ISM temperatures $< 250$K.

In order to model the atomic hydrogen column density correctly,
we take care to include the portion of the {\it{HST}}/STIS spectrum
exhibiting the Ly $\alpha$ feature into the modeling.  
In Figure 4 we show our fit to the
(atomic and molecular) hydrogen absorption lines to  
the {\it{FUSE}} spectrum of V794 Aql, as applied to our WD model fit
(namely, we multiplied the WD synthetic spectral model by the transmission
values obtained from the ISM model). 
This ISM model has zero metalicity, a temperature
of 150K, a turbulent velocity of 40km$~$s$^{-1}$, a molecular hydrogen column
density of $3 \times 10^{17}$cm$^{-2}$ and an atomic hydrogen column density of 
$4.5 \times 10^{20}$cm$^{-2}$. 
The Ly $\beta$ feature around 1025\AA\ is obviously
far too sharp to be due to the disk, even with an inclination of only $20\deg$
this feature would be much broader. This absorption feature cannot be
accounted by the star itself neither and it is modeled here as part
of the ISM model. 

It is interesting to note that the hydrogen column density of V794 Aql 
that we obtain here is very similar to the hydrogen column density of V1432 Aql
\citep{ran05} which is located in the same region of the sky (at about 230 pc). 
V1432 Aql is much closer to V794 Aql than any of the nearest neighboring sources
found by the ISM hydrogen column density tool from the {\it{EUVE}}
Web site. Also in the same vicinity and at about the same distance 
($250 \pm 100$pc), UU Aql exhibits strong hydrogen 
absorption features in it {\it{FUSE}} spectrum \citep{sio07}.  
This seems to indicate that this region of sky in the constellation of
Aquila might have an ISM cloud in the foreground (say at $d< 200$pc or less).  
  
Next, we use the hydrogen column density we obtained to estimate 
the reddening of the system using the analytical
expression given by \citet{boh78}: 
\begin{equation} 
E(B-V) =  \frac{N(H\small{I} +H_2)} {5.8 \times 10^{21}atoms~cm^{-2}mag^{-1}}.  
\end{equation} 
We are aware that this expression
is not accurate, in that sens that it represents a mean extinction law
with a large deviation,  as the ratio of dust to hydrogen is not 
constant. For example, objects towards which the atomic column density is (say)
$ 4 \times 10^{20}$atoms cm$^{-2}$, have a E(B-V) value ranging from
0.03 to 0.2, almost a factor of 10. For hydrogen column densities 
higher than $\approx 10^{21}$atoms cm$^{-2}$,  E(B-V) varies only by   
a factor of 3 but the typical scatter is really of only about 30\%.  
 
Using our computed ISM value for the hydrogen column density 
$N(H_2)=3 \times 10^{17}$cm$^{-2}$ and 
$N(H\small{I})= 4.5 \times 10^{20}$cm$^{-2}$ in equation (1), 
we obtain $E(B-V)=0.08$.  
The actual reddening toward V794 Aql could be smaller or larger 
than 0.08 as explained in the previous paragraph. However, from the 
visual inspection of m(AB) of the {\it{IUE}} spectra of V794 Aql we decide 
to model the spectra of V794 Aql for 2 values of the reddening,
namely E(B-V)=0.1 and E(B-V)=0.2.  
 
\subsection{The {\it{HST}}/STIS Spectrum}

An {\it{HST}}/STIS snapshot spectrum of V794 Aql was obtained on
August 28 2003, totaling 830 sec of exposure time. The spectrum
was taken with the 0.2x0.2 aperture in ACCUM (accumulation) mode
using the G140L grating, covering the wavelength range from
1150 to 1710\AA , with a wavelength binning of $\approx 0.6$\AA . 
The use of G140L provides a higher
resolution spectrum than {\it{IUE}}, but not as high as when using
E140M. The spectrum was taken in a relatively high state, similar
to the state during which the {\it{FUSE}} spectrum was obtained. 
The spectrum exhibits some broad emission lines, mainly
$C_{III}$ (1175\AA ), $N_{V}$ (1240\AA ), $Si_{IV}$ (1400\AA ), 
$C_{IV}$ (1550\AA ) and $He_{II}$ (1640\AA ). 
Other possible lines are also annotated in Figure 1. 
Also very pronounced is the Ly $\alpha$ absorption feature
(around 1215\AA ). 
On the sides of the Ly$\alpha$ absorption feature 
there is some emission from $N_I$ on the left and $Si_{II}$ on the right.  
The Ly$\alpha$ absorption feature is too sharp to be
that of the WD and/or the accretion disk of V794 Aql, 
however, it agrees very well with the hydrogen column density
we found, except that it should be saturated, 
as the transmission values of the ISM model 
are null there due the high column density of atomic hydrogen. 
The only reason the Ly$\alpha$ is not saturated in
the observed STIS spectrum of V794 Aql is  because  
the spectral resolution ($2 \times 0.6= 1.2$\AA ) 
of the G140L grating combined with the short exposures required
by snapshots often does
not resolve the sharp geocoronal $H_{I}$ emission, seen in many
STIS G140L exposures of some systems. 
For example the (MAST) archived STIS snapshot of VY Scl (with 
E(B-V)=0.06 \citep{bru94}) has a sharp
Ly$\alpha$ absorption feature that is not saturated,    
and the STIS snapshots of SS Aur and V442 Cen (with 
E(B-v)=0.08 and 0.15, respectively; \citep{bru94})
show clearly a sharp emission peak in the bottom of their Ly$\alpha$
profile. All these systems with a reddening value as large as that of
V794 Aql have most probably an atomic hydrogen column density larger
than $10^{20}$cm$^{-2}$ and should therefore all have a saturated Ly$\alpha$
absorption feature (for any line of sight
with $N(HI)>10^{14}$cm$^{-2}$ the Ly$\alpha$ 
absorption feature should be saturated). 

\section{Synthetic Spectral Modeling} 

\subsection{Preparation of the Spectra} 

Not enough light-curve data points were available from 
the American Association of Variable
Star Observers (AAVSO) to determine whether V794 Aql was observed in a high 
or low state. However, a comparison of the 
{\it{FUSE}} spectrum with the {\it{IUE}} 
archival spectra and the {\it{HST}}/STIS snapshot 
(see Figure 1) helps us to assess
the state in which the system was at the time each observations was made.  
From that figure it is clear that V794 Aql was observed in about
four different states. 

The lowest {\it{IUE}} spectrum has a flux about 5 times
lower than the {\it{FUSE}} and STIS spectra. 
This {\it{IUE}} spectrum, however, was
itself not obtained during the lowest state($B=20$). It seems therefore quite
clear that the {\it{FUSE}} and STIS spectra 
need mainly to be modeled with a disk model and the contribution
of the WD might be small. However, because 
of its inclination ($i \approx 39 \pm 17 \deg$)
the contribution of the heated WD during outburst might be non-negligible.  
Though there is a large error bar on the mass of the WD
(0.88 $\pm$ 0.4 $M_{\odot}$), 
we deliberately decided to use only one mass in our modeling
($M=0.9 M_{\odot}$) 
to limit the number of
unknown parameters in the system: the distance, the inclination
$i=39 \pm 17 \deg$, the reddening.   
We deredden the spectra assuming a reddening of E(B-V)=0.1 and 0.2.

We prepared the {\it{FUSE}} and
{\it{HST}} spectra for fitting by masking regions containing emission
lines and artifacts. These regions are emphasized with a blue color
in Figure 3.  
These regions of the spectrum were not included in the fitting.
A comparison of the {\it{HST}}/STIS spectrum with the {\it{FUSE}}
spectrum in the wavelength overlap region reveals that the flux levels
match thus enabling us to carry out model fits over a substantially
broader wavelength range. 

\subsection{The Synthetic Spectra} 

The high-gravity white dwarf atmosphere model spectra are generated 
assuming solar abundances (unless otherwise specified) 
using TLUSTY200 \citep{hub88}, SYNSPEC48 and ROTIN4 \citep{hub95}. 
TLUSTY generates numerical models of stellar atmospheres for a given 
surface gravity and effective temperature. SYNSPEC is then used
to generate a synthetic spectrum for each particular stellar atmosphere model. 
The routine ROTIN is used last to perform a rotational and/or
instrumental convolution of the synthetic spectrum obtained from SYNSPEC. 
For the input  temperature  in TLUSTY, 
we chose T ranging from 15,000K to 55,000K by increments of 5,000K
at first, and when a best fit is found a refined fitting is carried out
by changing the temperature in increments of 1,000K.
We chose a value of $\log{g}$ to match the mass of the WD in V794 Aql, namely 
$\log{g}=8.5$. 
We also varied the stellar rotational velocity 
$V_{rot} \sin{i}$ from $100$km$~$s$^{-1}$
to $600$km$~$s$^{-1}$ in steps of $100$km$~$s$^{-1}$. We have the possibility
to change the abundance of elements, however, we first ran a model with solar 
composition and only changed the Si and C abundances in two cases, to assess
how it affects the results. 

For the synthetic accretion disk models, 
we used the latest accretion disk models from 
the optically thick disk model grid of \citet{wad98}.
The range of disk model parameters in the grid of \citet{wad98}
varies as follows: WD mass (in solar
masses) values of 0.35, 0.55, 0.80, 1.03, and 1.21; orbital inclination (in
degrees) of 18, 41, 60, 75, and 81. The accretion rate ranges from
$10^{-10.5} M_{\odot}$yr$^{-1}$ to $10^{-8.0} M_{\odot}$yr$^{-1}$, varying in
increments of 0.5 in log \.{M}. 
For the present work we select from this grid 
the models with parameters consistent with Table 1, namely: 
WD masses 0.80 and 1.03$M_{\odot}$, an inclination angle of 18, 41 and 60$\deg$
and the entire range of mass accretion rates from the grid of \citet{wad98}.   

For combined accretion disk and stellar WD atmosphere models, the disk 
flux is divided by 100 to normalize it at 1000pc to match the WD flux, 
therefore giving explicitly the relative flux contributions of each
component. Then both fluxes are added for comparison with the observed flux. 
The best two-component models are not especially 
a combination of the best WD models with the best disk models. 
The main aim here is to find a fit that is better (lower $\chi^2_{\nu}$) 
than both the WD model fit and the accretion disk model fit. 

Before finding the best fit, all the synthetic spectral models 
(WD, accretion disk, and WD+accretion disk) 
are multiplied by the transmission values of 
the ISM model presented in section 2.2.

\subsection{The $\chi^2$ Minimization Routine and the Best Fit}  

In order to find the best model, 
we use FIT - a $\chi^2$ minimization routine (see e.g. \citet{pre92}).
For each model, we computed $\chi^{2}_{\nu}$ and scale factor values.
$\chi^2_{\nu}$ is known as the ``reduced'' chi-squared, namely $\chi^2$
per number of degrees of freedom. In the present case the number
of degrees of freedom is the number of wavelength bins of the observed
spectrum taken into account in the fitting. Therefore, when we exclude
(i.e. mask) regions of the spectrum (such as broad emission lines and features
such as air glow) in preparation for the fitting, we reduce the number of
degrees of freedom. And when we merge the {\it{FUSE}} spectrum together with the
{\it{HST}}/STIS spectrum we actually increase the number of degrees of freedom. 
Since {\it{FUSE}} and STIS are binned differently and have different errors, 
one cannot compare the least chi square best fit model obtained by fitting
the {\it{FUSE}} spectrum alone with the least chi square best fit model 
obtained when fitting the STIS spectrum alone or the 
combined {\it{FUSE}}+STIS spectrum. These fits 
cannot be compared on the basis of their $\chi^2_{\nu}$ values, as the
same synthetic model will lead to different $\chi^2_{\nu}$ values 
when fitting the {\it{FUSE}} spectrum alone, 
the STIS spectrum alone, or the combined {\it{FUSE}}+STIS spectrum.   
So the analysis of the {\it{FUSE}}, STIS  
and the combined {\it{FUSE}}+STIS spectra are carried out separately.  

For each single value of $\log{g}$ there exists only one value of 
$R_{wd}$ and $M_{wd}$, since we use the mass radius relation for WDs  
(e.g. \citet{ham61} or \citet{woo90} for different composition and
non-zero temperature WDs). As a consequence, when fitting the theoretical flux
to the observed flux, the distance to the system is obtained as an output
parameter when scaling the fluxes. When the WD mass and distance of a
system are known, the fitting technique leads to only one precise value of
the WD temperature. In the present case, since neither the mass nor the distance
are known, the best fit model consists actually 
of a whole domain in the $\log{g}$-$T_{eff}$ 
plane. Because of that we fix the WD mass to $0.9M_{\odot}$ (for the WD models)
and use the $M_{wd}=0.80, 1.03M_{\odot}$ models from the grid of accretion 
disk models. We then chose the lowest $\chi^2$ model (best fit) and obtain as
output the WD temperature (or the mass accretion rate) 
and the distance to the system. Since the 
reddening is also not known accurately, the choice $M=0.9M_{\odot}$
helps us limit the number of solutions to a single table.    

\section{Results and Discussion} 

We first tried a single white dwarf model, then an accretion disk alone, 
and then a combination of both for the {\it{FUSE}} spectrum alone, then for
the {\it{HST}}/STIS spectrum alone and then for 
the combined {\it{FUSE}} + {\it{HST/}}STIS
spectrum. . 
We note that if we do not deredden the observed  spectra, they
are basically impossible to fit with our synthetic spectral
models.   

\subsection{The {\it{FUSE}} Spectrum} 

The best solar composition WD model fit we found 
to the dereddened {\it{FUSE}} spectrum of V794 Aql assuming E(B-V)=0.1  
has an effective temperature of 44,000K, a rotation rate of 200km$~$sec$^{-1}$
and a distance of 301pc. This model is shown in Figure 4 
and listed in Table 5.  
The WD model has some absorption features (around 1120-1130\AA ) not
present at all in the observed spectrum. In order to try and improve the fit, 
we decrease the abundances of C and Si (responsible for these absorption
feature) to 0.01 their solar value and kept all the other abundances 
solar. This provides a 10 percent reduction in the $\chi^2_{\nu}$ value
(Table 5).   

Next we ran the fitting subroutine on the grid of accretion disk models
and found that the fitting is slightly better for the disk than
for the WD. 
The best fit accretion disk models are also listed in Table 5 for the
$M=0.8M_{\odot}$ and $M=1.03M_{\odot}$ cases.  
The mass accretion rate we obtained is rather large 
($\dot{M}=10^{-9}-10^{-8.5}M_{\odot}$yr$^{-1}$) 
and the distance is much larger than for a WD model, namely $d=616-834$pc. 
In figure 5 we show the best fit accretion disk model 
with $M=1.03M_{\odot}$ and $\dot{M}=10^{-9}M_{\odot}$yr$^{-1}$. 
The inclination
angle we obtained is 60 degrees. In fact for all the disk models we
found in the present work, 
the best fit is obtained for the higher inclination. This comes
from the fact that the slope of the continuum is better matched by a rather
flat continuum, which is obtained naturally with a higher inclination.   
A flatter continuum can also be obtained by dereddening the spectrum 
assuming a higher E(B-V) value, which is what we did next.

We ran our fitting subroutine for the WD and accretion disk models 
assuming now E(B-V)=0.2, and dereddening the {\it{FUSE}} spectrum of V794 Aql 
accordingly. We found that the $\chi^2_{\nu}$ value decreases by about
20 percent for all the models. However, now the best WD model has a
temperature of 51,000K (since the spectrum is now "bluer") and a
distance of 200pc. The best WD model fit assuming low C and Si 
abundances is presented in Figure 6 (see also Table 5). 
For the best accretion disk model we found a
slightly higher mass accretion rate 
($\dot{M}=10^{-8.5}-10^{-8.0}M_{\odot}$yr$^{-1}$) than for the E(B-V)=0.1
case, and a distance $d=585-791$pc. Here again the
best fit is for the accretion disk models. The $1.03M_{\odot}$
model (Figure 7) only marginally better than the $0.8M_{\odot}$ model
(all the models are listed in Table 5). 

Next we ran two-component (disk+WD) model fits to find whether
the fitting can be improved, but we found that the least $\chi^2_{\nu}$
obtained for the two-component is never smaller than 
the least $\chi^2_{\nu}$ obtained for the disk alone. 

From the {\it{FUSE}} spectrum alone, we find that the best fit model
is that of an accretion disk with a high accretion rate, in
agreement with the fact that V794 Aql was caught in a rather high
state. The distance inferred from the modeling is pretty large 
$d\approx 700 \pm 100$pc, consistent with the assumption of a large
reddening value (0.2) and with the rather low {\it{FUSE}} 
flux for a high state.   
 
\subsection{The {\it{HST}}/STIS Spectrum}  

Next we carried out exactly the same analysis but for the {\it{HST}}/STIS
spectrum of V794 Aql, assuming both E(B-V)=0.1 and E(B-V)=0.2.
For easy comparison all these results are also listed in Table 5. 
The main  difference with the {\it{FUSE}} results is that the temperature of
the best WD model is 6,000K higher for the E(B-V)=0.1 case 
and 4,000K higher for the E(B-V)=0.2 case. The distance for the
WD models is about the same, namely d=200-300pc. 
For the E(B-V)=0.1 case, the best accretion disk model (Figure 8) has a slightly
smaller mass accretion rate  
($\dot{M}=10^{-9.0}-10^{-9.5}M_{\odot}$yr$^{-1}$) than for the {\it{FUSE}}
best fit which leads to 
about half the distance obtained for the {\it{FUSE}} spectrum with the
same reddening value. For the E(B-V)=0.2 case we obtained exactly the
same best accretion disk solution as for the {\it{FUSE}} spectrum. 
The best fit WD model (for E(B-V)=0.2) is presented in Figure 9 and the 
best fit accretion disk model (for E(B-V)=0.2) is presented in Figure 10. 
Note that for all the STIS spectral fits  
the $\chi^2_{\nu}$ values for the disk models 
are significantly smaller than for the best WD models, namely:  
$\chi^2_{\nu}$=13.25 versus  21.65 for the E(B-V)=0.1 case, and 
$\chi^2_{\nu}$=9.63 versus  13.65 for the E(B-V)=0.2 case.  
This is an indication that from the STIS spectrum 
alone, the accretion disk
is unambiguously the best model fit, and that E(B-V)=0.2 might be the correct
value for the reddening.  
Here too the two-component (disk+WD) model did not lead to a better fit. 

\subsection{The Combined {\it{FUSE+HST}}/STIS Spectrum}  

In order to combine the {\it{FUSE}} spectrum with the STIS spectrum 
we check how well their fluxes match in the overlap region 
between about 1150\AA and 1180\AA .  At very short wavelengths STIS is 
very noisy, and the longer wavelengths of {\it{FUSE}} are represented 
only by one channel (because of the 'worm') and are, therefore, less 
reliable too. Consequently, we are left only with the $C_{III}$ emission 
region and its immediate vicinity. We find that the spectra have the
same flux level there and can therefore be combined together.   
 
Again we ran our fitting subroutine, but now for the combined {\it{FUSE}}+STIS
spectrum of V794 Aql. This time we found that the temperature for the
best WD model fit is much lower than for the individual spectra,
namely 30,000K for the E(B-V)=0.1 case and T=45,000K for the
E(B-V)=0.2 case (shown in Figure 12) with a distance of only d=145-176pc.  
For the E(B-V)=0.1 case (Figure 11), we found the same accretion disk solution as for the
STIS spectrum, while for the E(B-V)=0.2 case (Figure 13) we found the same accretion
disk solution as for the {\it{FUSE}} and STIS spectra alone. Again the lowest  
$\chi^2_{\nu}$ values were obtained for the accretion disk models.
Again the  E(B-V)=0.2  solutions were better than the E(B-V)=0.1
solutions, and here too the two-component (disk+WD) model did 
not lead to a better fit. 

\subsection{The {\it{IUE}} Spectrum in a Low State}   

Since the {\it{FUSE}} and  {\it{HST}}/STIS spectra were obtained in a 
relatively high state, we could not get much information on the
WD itself. In order to try and gain some knowledge of the WD
directly we decided to model the {\it{IUE}} spectrum SWP 15266
of V794 Aql in a relatively low state. In this state we expect 
to see mainly the WD and therefore we model the spectrum with
a $0.8M_{\odot}$ WD to agree with the grid of accretion disk 
models. Since the spectrum is pretty noisy, more than
one temperature is obtained and to limit the number of solution we
fixed the distance to d$\approx$200pc and d$\approx$600pc and dereddened 
the spectrum assuming E(B-V)=0.20. 

The best fit model for the shortest 
distance consists of a WD alone with T=30,000K, with  $\chi^2_{\nu}$=
5.17 and a distance of 244pc. 

For the larger distance the best fist consists of a WD plus and accretion 
disk model (Figure 14). The WD has T=47,000K and the disk has 
$\dot{M}=10^{-9.5}M_{\odot}$yr$^{-1}$, i=41deg. This best model has
a distance of 643pc and $\chi^2_{\nu}$=4.52. In this model the
WD contributes 56 percent of the flux and the accretion disk
contributes the remaining 44 percent.  
If this temperature is correct, then we have to reject the WD
models from Table 5 with $T<47,000K$ for E(B-V)=0.20; this would
be the WD model for the combined {\it{FUSE}}+STIS spectrum for
(E(B-V)=0.20) with a temperature of only 45,000K, as this would
be inconsistent, namely the system during the higher state cannot
have a temperature lower than during quiescence (implying $T>47,000K$).  

\section{Summary}

We have analyzed the FUV spectra of V794 Aql observed in 
a relatively high state with {\it{FUSE}} and {\it{HST}}/STIS
and found evidence for the presence of a hot accretion disk
accreting at a high rate. 
The spectra exhibit some broad emission
lines ($O_{VI}$, $C_{III}$, $N_V$, $Si_{IV}$, $C_{IV}$
and $He_{II}$), a clear sign of a hot gas where a high rotational
velocity in a disk might be responsible for the broadening of
these lines. 
We were limited in our spectral analysis by the large
number of unknown parameters (distance, WD mass, inclination, reddening,
WD temperature,..) and decided to limit the search of solution
assuming $M_{wd}=0.9M_{\odot}$.   
For the {\it{FUSE}}, STIS and {\it{FUSE}}+STIS spectra of V794 Aql, 
we found that the accretion disk model provides a better fit to the
observed spectra, which is not surprising since the system was observed
in a relatively high state when the contribution of the disk is dominant. 
The best fit model had a 
mass accretion rate of $10^{-8.5}-10^{-8.0}M_{\odot}$yr$^{-1}$ and  
an inclination of 60 degrees, assuming a mass $M=0.9M_{\odot}$. 
The distance we obtained was $d =690 \pm 105$pc and in
all the cases the least $\chi^2_{\nu}$ was obtained  for
E(B-V)=0.2. 
We also found that V794 Aql is moderately  
affected by the ISM with a molecular hydrogen column density of
$3 \times 10^{17}$cm$^{-2}$ 
and an atomic hydrogen column density of 
$4.5 \times 10^{20}$cm$^{-2}$.

\acknowledgments 

PG wishes to thank the Space Telescope Science Institute
for its kind hospitality. 
This research was partly based on observations made with the 
NASA-CNES-CSA Far Ultraviolet Spectroscopic Explorer. 
{\it{FUSE}} is operated for NASA by the Johns Hopkins University under
NASA contract NAS5-32985. 
Funding was provided by  NASA {\it{FUSE}} (cycle 4) grant NNG04GL45G 
to Villanova University (P.Godon), NASA {\it{FUSE grant}} NNG04GC97G 
to the University of Washington (P.Szkody), and NASA {\it{FUSE}} grant 
NNG04GL18G (P.E.Barrett). 
Additional support for this work was provided by NASA through
grant number HST-AR-10657.01-A (HST Cycle 14 Archival) 
to Villanova University (P.Godon) and grant number HST-GO-09724 
to the University of Washington (P.Szkody), 
from the Space Telescope Science Institute,
which is operated by the Association of Universitie for 
Research in Astronomy, Incorporated, under NASA contract NAS5-26555.

\clearpage 

\begin{center}
\begin{tabular}{ll}
\hline
\multicolumn{2}{c}{Table 1:
V794 Aql System Parameters}\\ 
\hline
\hline
Subtype:               & VY Sculptoris    \\
P$_{orb}$(d):          & 0.1533$^a$       \\
inclination $i$ (deg): & $39 \pm 17$      \\
M$_{1}(M_{\odot}$):    & $0.88 \pm 0.39$  \\
V$_{max}$:             & 14.0          \\
V$_{min}$:             & 20.2$^b$        \\
\hline
\end{tabular}
\end{center}
\begin{center}	
\small (a)\citet{hon98} (b) the visual magnitude of V794 Aql
usually ranges between $\approx 14$ and $18$ \citep{hon98}
but it has also been observed in an extremely low state
where the magnitude drops to $\approx 20$ \citep{hon85}. 
\end{center}

\hspace{-2.cm} 
\begin{tabular}{clcccl}
\multicolumn{6}{c}
{Table 2: Observing Log for V794 Aql}\\ \\  
\hline  \\ 
Telescope& dataset      &    Date       & t$_{\rm exp}$(s) & Aperture & Additional Characteristics   \\ \\  
\hline \\ 
IUE      & SWP 15266    &   10/14/81    & 7200             & Large    & Disp.=low, Contin= 54 cts/s, Bckgr=28 cts/s  \\
IUE      & SWP 28501    &   06/16/86    & 23400            & Large    & Disp.=low, Contin=118 cts/s, Bckgr=73 cts/s \\ 
IUE      & SWP 50754    &   05/13/94    & 7198             & Large    & Disp.=low, Contin=166 cts/s, Bckgr=96 cts/s \\ 
IUE      & SWP 14708$^1$&   08/10/81    & 3600             & Large    & Disp.=low, Contin=0 cts/s, Bckgr=23 cts/s  \\ 
IUE      & LWR 11782    &   15/10/81    & 3600             & Large    & Disp.=low, Contin=72 cts/s, Bckgr34 cts/s \\ 
FUSE     & D1440101     &   05/13/04    & 13,333           & LWRS     & mode=TTAG, Cent.Wave.=959.998 \AA\      \\  
HST/STIS & O8MZ66010    &   08/28/03    & 830              & 0.2x0.2  & mode=ACCUM, Filter/Grating=G140L       \\    \\ 
\hline  \\  
\end{tabular}
(1) SWP 14708 has a flux level of the order of $10^{-15}$erg$~$s$^{-1}$cm$^{-2}$\AA$^{-1}$ 
and cannot be used.

\clearpage

\begin{center}
\hspace{-2.cm} 
\begin{tabular}{lccl}
\multicolumn{4}{c}
{Table 3: FUSE lines}\\ \\  
\hline 
Line            & Wavelength & Absorption$^{(1)}$  & Origin$^{(2)}$     \\ 
Identification  & \AA\       & Emission    &          \\ 
\hline 
$S_{VI}$         & 944.50    &   e       &        s     \\ 
$S_{VI}$         & 933.50    &   e       &        s     \\ 
$N_{I}$          & 964.63    &   a       &       ism    \\ 
$C_{III}$        & 977.02    &   e       &       c,s     \\ 
$Si_{II}$        & 1020.70   &   a       &       ism,s     \\ 
$O_{VI}$         & 1031.93   &   e       &       c,s     \\ 
$O_{VI}$         & 1037.62   &   e       &       c,s     \\ 
$Ar_{I}$         & 1048.20   &   a       &       ism    \\ 
$Ar_{I}$         & 1066.66   &   a       &       ism    \\ 
$S_{IV}$         & 1073.52   &   a       &       s         \\ 
$N_{II}$         & 1083.99   &   a,e     &       c,ism    \\ 
            & 1084.56   &   a,e     &       c,ism    \\ 
            & 1084.58   &   a,e     &       c,ism    \\ 
            & 1085.53   &   a,e     &       c,ism    \\ 
            & 1085.55   &   a,e     &       c,ism    \\ 
            & 1085.70   &   a,e     &       c,ism    \\ 
$S_{I}$          & 1096.60   &   a       &       s     \\ 
$Fe_{II}$        & 1096.88   &   a       &       ism     \\ 
$Fe_{II}$        & 1125.45   &   a       &       ism     \\ 
$N_{I}$          & 1134.16   &   a       &       ism        \\ 
            & 1134.42   &   a       &       ism        \\ 
            & 1134.98   &   a       &       ism        \\ 
$Fe_{II}$        & 1144.94   &   a       &       ism    \\ 
$P_{II}$         & 1152.82   &   a       &       ism    \\ 
$He_{I}$         & 1168.61   &   e       &       c       \\ 
$S_{I}$          & 1172.55   &   a       &       s    \\ 
\hline  \\    
\end{tabular}
\end{center} 
(1): e=emission,  a=absorption.\\   
(2): c=contaminated by airglow, geocoronal or heliocoronal emission, 
s=system,  ism=interstellar medium  \\ 
The contamination is from geocoronal lines mainly: 
$H_{I}$, $O_{I}$, $N_{I}$ and $N_{II}$, and scattered solar light, such as  
the sharp $C_{III}$ (977\AA ) and $O_{VI}$ (1032\AA ) 
which are mainly seen here in the SiC channels. 

\begin{center}
\hspace{-2.cm} 
\begin{tabular}{llll}
\multicolumn{4}{c}
{Table 4: Molecular Hydrogen Absorption Lines}\\ \\  
\hline 
Line            & Wavelength  &   Line            & Wavelength     \\ 
Identification  & (\AA\ )     &   Identification  & (\AA\ )        \\ 
\hline 
$H_2$ blend    & 946-948$~~~~~~~~~~~~~~~~~~~$ &    L6R1        & 1024.99           \\ 
L13R1       & 955.06          &    L5R0        & 1036.55           \\ 
L13P1       & 955.71          &    L5R1        & 1037.15           \\ 
L12R0       & 962.98          &    L5P1        & 1038.16           \\ 
L12R1       & 963.61          &    L5R2        & 1038.69           \\ 
$H_2$ blend    & 965.00          &    L5P2        & 1040.37           \\ 
W2Q1        & 966.09          &    L5R3        & 1041.16           \\ 
W2Q2        & 967.28          &    L5P3        & 1043.50           \\ 
L10R0       & 981.44          &    L4R0        & 1049.37           \\ 
L10R1       & 982.07          &    L4R1        & 1049.96           \\ 
L10P1       & 982.84          &    L4P1        & 1051.03           \\ 
L10R2       & 983.59          &    L4R2        & 1051.50           \\ 
L10P2       & 984.86          &    L4P2        & 1053.28           \\ 
W1R0+W1R1   & 985.60          &    L4R3        & 1053.98           \\ 
W1R1        & 986.80          &    L4P3        & 1056.47           \\ 
W1R2        & 987.97          &    L3R0        & 1062.88           \\ 
L9R0        & 991.38          &    L3R1        & 1063.46           \\ 
L9R1        & 992.01          &    L3P1        & 1064.61           \\ 
L8R0        & 1001.82         &    L3R2        & 1065.00           \\ 
L8R1        & 1002.45         &    L3P2        & 1066.90           \\ 
L8P1        & 1003.29         &    L3R3        & 1067.48           \\ 
L8R2        & 1003.98         &    L3P3        & 1070.14           \\ 
W0R0+W0R1   & 1008.50         &    L2R0        & 1077.14           \\ 
W0R2        & 1009.02         &    L2R1        & 1077.70           \\ 
W0R1        & 1009.77         &    L2P1        & 1078.93           \\ 
W0R2        & 1010.94         &    L2R2        & 1079.23           \\ 
W0P2        & 1012.17         &    L1R0        & 1092.20            \\ 
L7R0        & 1012.70         &    L1R1        & 1092.73            \\ 
L7R1        & 1013.44         &    L1P1        & 1094.05            \\ 
L7P1+W0P3   & 1014.50         &    L1P3        & 1099.79            \\ 
L7R2        & 1014.97         &    L0R0        & 1108.13            \\ 
L7P2        & 1016.46         &    L0R1        & 1108.63            \\ 
L7R3        & 1017.42         &    L0P1        & 1110.06            \\ 
L6R0        & 1024.37         &    L0R2        & 1110.12            \\ 
\hline  \\    
\end{tabular}
\end{center} 

\clearpage 

\begin{center}
\begin{tabular}{ccccccccc}
\multicolumn{9}{c}{Table 5: Model Fits to the spectrum of V794 Aql}\\
\hline
$M_{wd}$ & $T_{wd}$ & Log($\dot{M}$) & $i$ &$\chi^2_{\nu}$& $d$&WD/disk&E(B-V)& Spectrum \\
($M_{\odot}$)&(1000K) &($M_{\odot}$yr$^{-1}$)&(deg)&   &(pc) &     &  &              \\
\hline
0.90         & 44       &  -       &  -  & 0.02959      & 301 & WD      & 0.10 & FUSE  \\
0.90         & 44       &  -       &  -  & 0.02767      & 304 & WD$^1$  & 0.10 & FUSE  \\
0.80         & -        & -8.5     & 60  & 0.02651      & 834 & disk    & 0.10 & FUSE  \\ 
1.03         & -        & -9.0     & 60  & 0.02584      & 616 & disk    & 0.10 & FUSE  \\ 
\hline 
0.90         & 51       &  -       &  -  & 0.02355      & 199 & WD      & 0.20 & FUSE  \\
0.90         & 51       &  -       &  -  & 0.02257      & 201 & WD$^1$  & 0.20 & FUSE  \\
0.80         & -        & -8.0     & 60  & 0.02206      & 791 & disk    & 0.20 & FUSE  \\ 
1.03         & -        & -8.5     & 60  & 0.02177      & 585 & disk    & 0.20 & FUSE  \\ 
\hline
 0.90         & 50       & -       & -   & 21.65        & 281 & WD      & 0.10 & STIS  \\
 0.80         & -        & -9.0    & 60  & 13.70        & 420 & disk    & 0.10 & STIS  \\ 
 1.03         & -        & -9.5    & 60  & 13.25        & 311 & disk    & 0.10 & STIS  \\ 
\hline
 0.90         & 55       & -       & -   & 13.65        & 201 & WD      & 0.20 & STIS  \\
 0.80         & -        & -8.0    & 60  &  9.63        & 794 & disk    & 0.20 & STIS  \\ 
 1.03         & -        & -8.5    & 60  &  9.82        & 589 & disk    & 0.20 & STIS  \\ 
\hline
 0.90         & 30       &  -      &       & 10.9    &  145  & WD      & 0.10  & FUSE+STIS \\  
 0.80         &   -      & -9.0    & 60    & 5.22    &  451  & disk    & 0.10  & FUSE+STIS \\  
 1.03         &   -      & -9.5    & 60    & 4.97    &  333  & disk    & 0.10  & FUSE+STIS \\  
\hline
 0.90         & 45       &  -      &       & 4.47    &  176  & WD      & 0.20  & FUSE+STIS \\  
 0.80         &   -      & -8.0    & 60    & 3.10    &  791  & disk    & 0.20  & FUSE+STIS \\  
 1.03         &   -      & -8.5    & 60    & 3.22    &  585  & disk    & 0.20  & FUSE+STIS \\  
\hline
 0.80         & 47       & -9.5    & 41    & 4.52    &  643  & 56/44   & 0.20  & IUE       \\  
 0.80         & 30       & -       & -     & 5.17    &  244  & WD      & 0.20  & IUE       \\  
\hline
\end{tabular}
\end{center}
\begin{center}	
\small (1) These WD model fits were slightly improved by decreasing C and Si abundances to 0.01 solar.  
\end{center}

\clearpage 

FIGURE CAPTIONS

Fig.1
- The {\it{FUSE}}, {\it{HST}}/STIS and {\it{IUE}} 
spectra of V794 Aql with line
identification in the STIS and {\it{IUE}} range. {\it{FUSE}} is in orange,
STIS is in black, the {\it{IUE}} SWP 50754 is in blue, SWP 28501 is in
red and SWP 15266 is in green. The {\it{FUSE}} spectrum has been binned
to 0.5\AA here. The flux (Y-axis) is given ergs$~$s$^{-1}$cm$^{-2}$\AA$^{-1}$
and the wavelength (X-axis) is given in \AA . 
The sharp emission lines in the {\it{FUSE}} spectrum 
and the strong Ly$\alpha$ emission in the {\it{IUE}} spectra are all
due to air glow. 
Line identification for {\it{FUSE}} is provided in Figure 3 and Tables 3 \& 4.

Fig.2
- The {\it{IUE}} SWP 28501 and LWR 11782 spectra of V794 Aql have been
combined together and binned at 20\AA\ following the procedure of 
\citet{ver87}. The AB magnitude is shown as a function of wavelengths
(\AA ) for the dereddened spectra assuming different value of 
E(B-V).  The broad absorption feature between 2000-2400\AA\ is clearly 
seen in the lower graph, a sign that E(B-V)$>$0. 

Fig.3
- The {\it{FUSE}} spectrum of V794 Aql with line identification. 
The flux (y-axis) is in erg$~$s$^{-1}$cm$^{-2}$\AA$^{-1}$ and the
wavelength (x-axis) is in \AA . The airglow (sharp) emission lines 
have been annotated with a circle with a cross; on top of the 
O{\small{VI}} and C{\small{III}} ($\lambda$977) broad emission from
the source, there is a contribution from helio-coronal emission;
The helium
and hydrogen emission from the airglow have been annotated under the 
x-axis for clarity. The short vertical lines without annotation
are the molecular hydrogen absorption lines due to the ISM.   
The nitrogen I and II are contaminated with airglow and ISM
absorption lines. The spectrum here has not yet been dereddened.

Fig.4
- The best synthetic spectral fit (solid black) using a WD stellar atmosphere 
model (with solar composition) to the dereddened {\it{FUSE}} spectrum
(in red and blue)  of V794 Aql assuming E(B-V)=0.1. 
The WD has a temperature $T_{eff}=44,000$K, 
a projected rotation rate of $V_{rot} \sin{i} =
200$km$~$s$^{-1}$, a mass $M=0.9 M_{\odot}$, a distance
of 301pc and a $\chi_{\nu}^2$=0.02959. The synthetic spectrum has
been multiplied by the transmission values of the ISM model to match
the ISM absorption feature. The strong emission lines (such as $O_{VI}$
and $C_{II}$), the ISM absorption features and the airglow lines have
all been masked and are shown here in blue. The dotted line shows the
synthetic spectrum without the ISM model. The fit is carried between the
solid black line and the red portions of the observed spectrum.  
Note in the synthetic spectrum the strong absorption features
between 1120 and 1130\AA , which are not present in the observed spectrum.  

Fig.5 
- The best synthetic spectral fit using a disk model to the 
dereddened {\it{FUSE}} spectrum of V794 Aql assuming E(B-V)=0.1. 
Using the grid of models from Wade 
and Hubeny, we find the best fit is for a WD with a mass $M=1.03M_{\odot}$,
a mass accretion rate is $\dot{M}= 10^{-9.0} M_{\odot}$yr$^{-1}$ an 
inclination of $i=60$deg and a distance of 616pc. 
The resulting reduced chi-square is $\chi_{\nu}^2$=0.02584.

Fig.6
- The best synthetic spectral fit (solid black) using a WD stellar atmosphere 
model (with low C and Si abundances ) to the dereddened {\it{FUSE}} spectrum
of V794 Aql assuming E(B-V)=0.2. 
The WD has a temperature $T_{eff}=51,000$K, 
a projected rotation rate of $V_{rot} \sin{i} =
200$km$~$s$^{-1}$, a mass $M=0.9 M_{\odot}$, a distance
of 201pc and a $\chi_{\nu}^2$=0.02257. 
Note in the synthetic spectrum that the absorption features
between 1120 and 1130\AA have been reduced by decreasing the
abundances of C and Si to better match the observed spectrum.  

Fig.7 
- The best synthetic spectral fit using a disk model to the 
dereddened {\it{FUSE}} spectrum of V794 Aql assuming E(B-V)=0.2. 
The best fit is for a WD with a mass $M=1.03M_{\odot}$,
a mass accretion rate is $\dot{M}= 10^{-8.5} M_{\odot}$yr$^{-1}$ an 
inclination of $i=60$deg and a distance of 585pc. 
The resulting reduced chi-square is $\chi_{\nu}^2$=0.02177.

Fig.8
- The best synthetic spectral fit (solid black) to the dereddened STIS 
spectrum (in red and blue) of V794 Aql assuming E(B-V)=0.1.
Here too the masked portions are shown in blue and the synthetic
spectrum without the ISM model is shown with the dashed line. 
The best fit consist of an accretion disk model 
with $\dot{M}= 10^{-9.5} M_{\odot}$yr$^{-1}$,  
around a WD with a mass $M=1.03M_{\odot}$,
an inclination of $i=60$deg and a distance of 311pc. 
The resulting reduced chi-square is $\chi_{\nu}^2$=13.25.

Fig.9
- The best synthetic spectral fit (solid black) using a WD stellar atmosphere 
model (solar abundances) to the dereddened STIS spectrum
of V794 Aql assuming E(B-V)=0.2. 
The WD has a temperature $T_{eff}=55,000$K, 
a projected rotation rate of $V_{rot} \sin{i} =
200$km$~$s$^{-1}$, a mass $M=0.9 M_{\odot}$, a distance
of 201pc and a $\chi_{\nu}^2$=13.65. 

Fig.10
- The best synthetic spectral fit to the dereddened STIS 
spectrum of V794 Aql assuming E(B-V)=0.2.
The best fit consist of an accretion disk model 
with $\dot{M}= 10^{-8.0} M_{\odot}$yr$^{-1}$,  
around a WD with a mass $M=0.80M_{\odot}$,
an inclination of $i=60$deg and a distance of 794pc. 
The resulting reduced chi-square is $\chi_{\nu}^2$=9.63.

Fig.11 
- The best synthetic spectral fit (solid black) to the dereddened 
combined ({\it{FUSE}}+STIS)  
spectrum (in red and blue) of V794 Aql assuming E(B-V)=0.1.
Here too the masked portions are shown in blue and the synthetic
spectrum without the ISM model is shown with the dashed line. 
The best fit consist of an accretion disk model 
with $\dot{M}= 10^{-9.5} M_{\odot}$yr$^{-1}$,  
around a WD with a mass $M=1.03M_{\odot}$,
an inclination of $i=60$deg and a distance of 333pc. 
The resulting reduced chi-square is $\chi_{\nu}^2$=4.97. 

Fig.12
- The best synthetic spectral fit using a WD stellar atmosphere 
model (solar abundances) to the dereddened combined ({\it{FUSE}}+STIS) 
spectrum of V794 Aql assuming E(B-V)=0.2. 
The WD has a temperature $T_{eff}=45,000$K, 
a projected rotation rate of $V_{rot} \sin{i} =
200$km$~$s$^{-1}$, a mass $M=0.9 M_{\odot}$, a distance
of 176pc and a $\chi_{\nu}^2$=4.47.

Fig.13 
- The best synthetic spectral fit (solid black) to the dereddened 
combined ({\it{FUSE}}+STIS)  
spectrum of V794 Aql assuming E(B-V)=0.2.
The best fit consist of an accretion disk model 
with $\dot{M}= 10^{-8.0} M_{\odot}$yr$^{-1}$,  
around a WD with a mass $M=0.80M_{\odot}$,
an inclination of $i=60$deg and a distance of 791pc. 
The resulting reduced chi-square is $\chi_{\nu}^2$=3.10. 

Fig.14 
The best fit model (in red) to the dereddened {\it{IUE}} spectrum SWP 15266
of V794 Aql (assuming E(B-V)=0.20) in a relatively low state. 
The best fit consist of a WD plus an accretion disk model.  
The WD has a mass $M=0.8M_{\odot}$ WD to agree with the grid of accretion 
disk models of Wade \& Hubeny. 
The best fit WD temperature for this model is 47,000K, the mass
accretion rate is $\dot{M}= 10^{-9.5} M_{\odot}$yr$^{-1}$,  
the inclination $i=41$deg and the distance 643pc.
The resulting reduced chi-square is $\chi_{\nu}^2$=4.52. 
In this model the WD (in blue) contributes to 56\% of the flux and the
disk (in green) contributes the remaining 44\%.

\clearpage 

\begin{figure}
\plotone{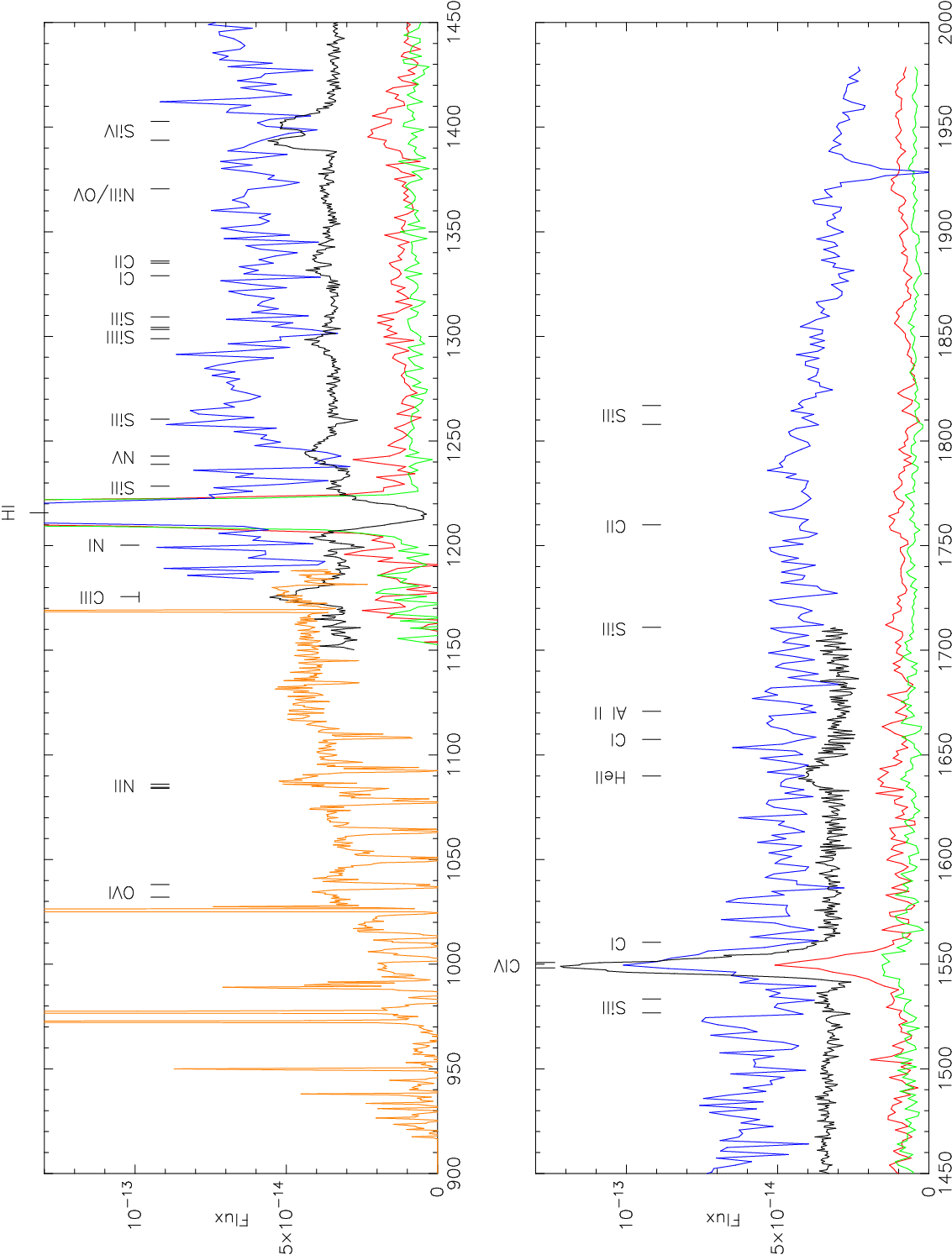}         
\end{figure} 

\clearpage 

\begin{figure}
\plotone{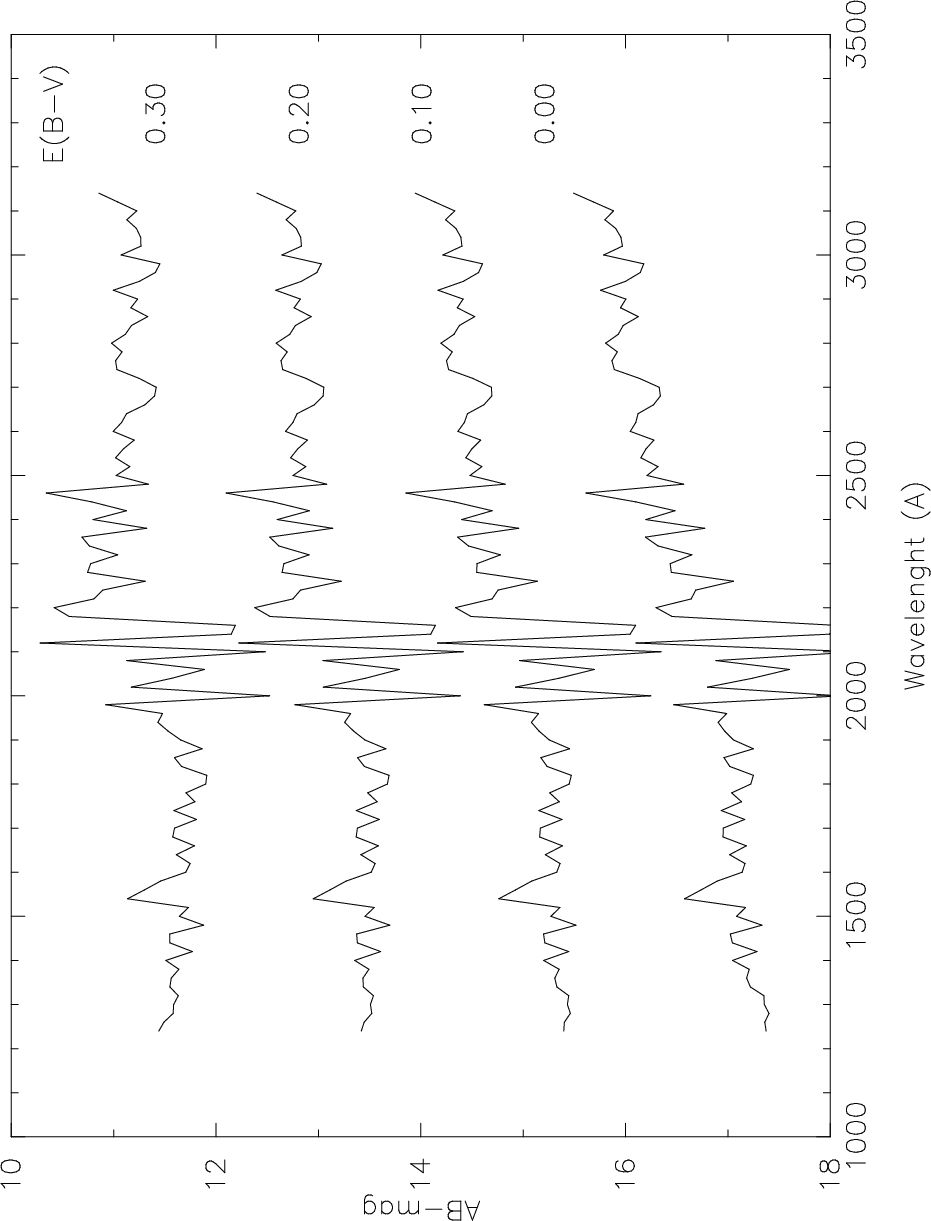}         
\end{figure} 

\clearpage 

\begin{figure}
\plotone{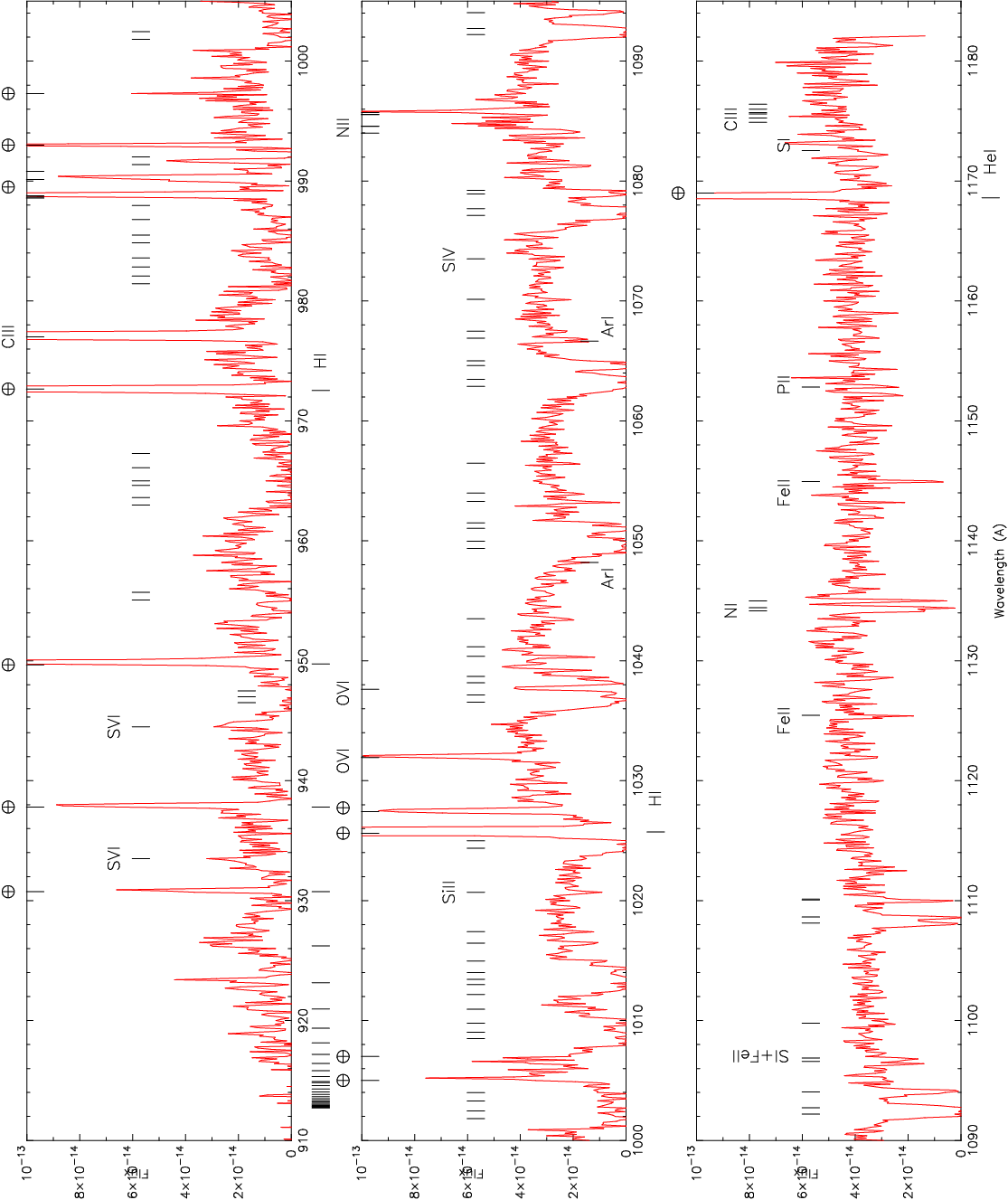}         
\end{figure} 

\clearpage 

\begin{figure}
\plotone{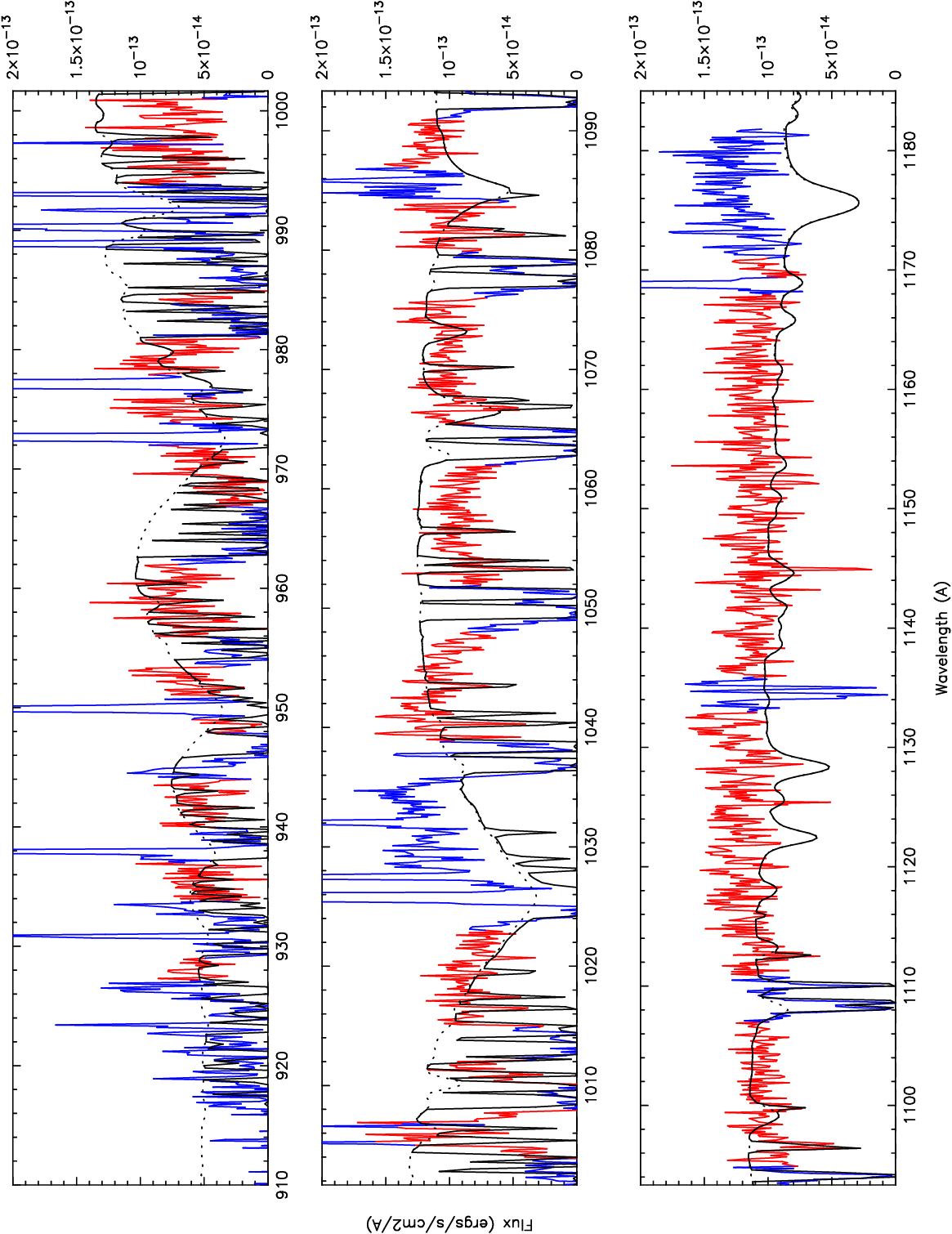}         
\end{figure} 
\clearpage 
     
\begin{figure}
\plotone{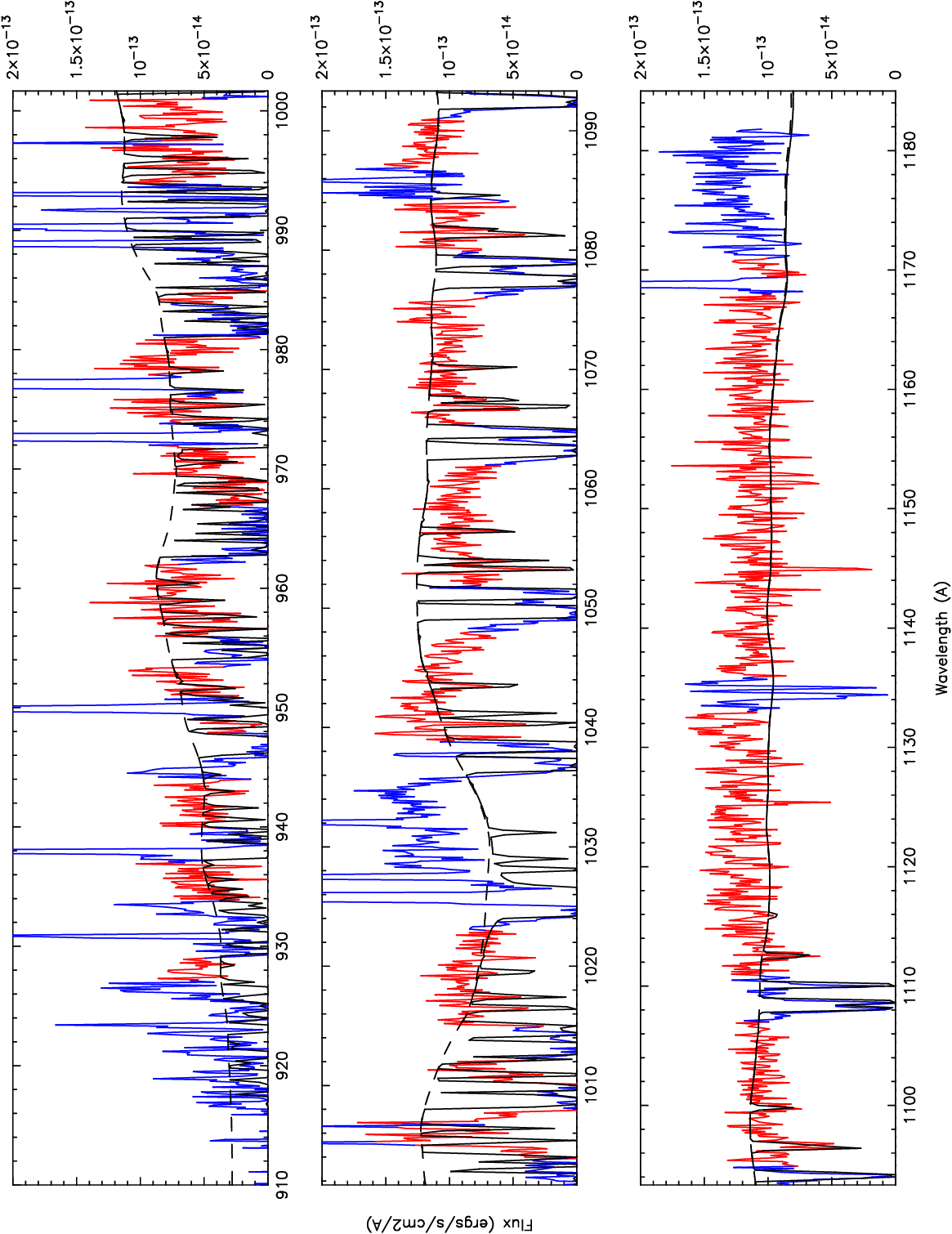}          
\end{figure} 
\clearpage 

\begin{figure}
\plotone{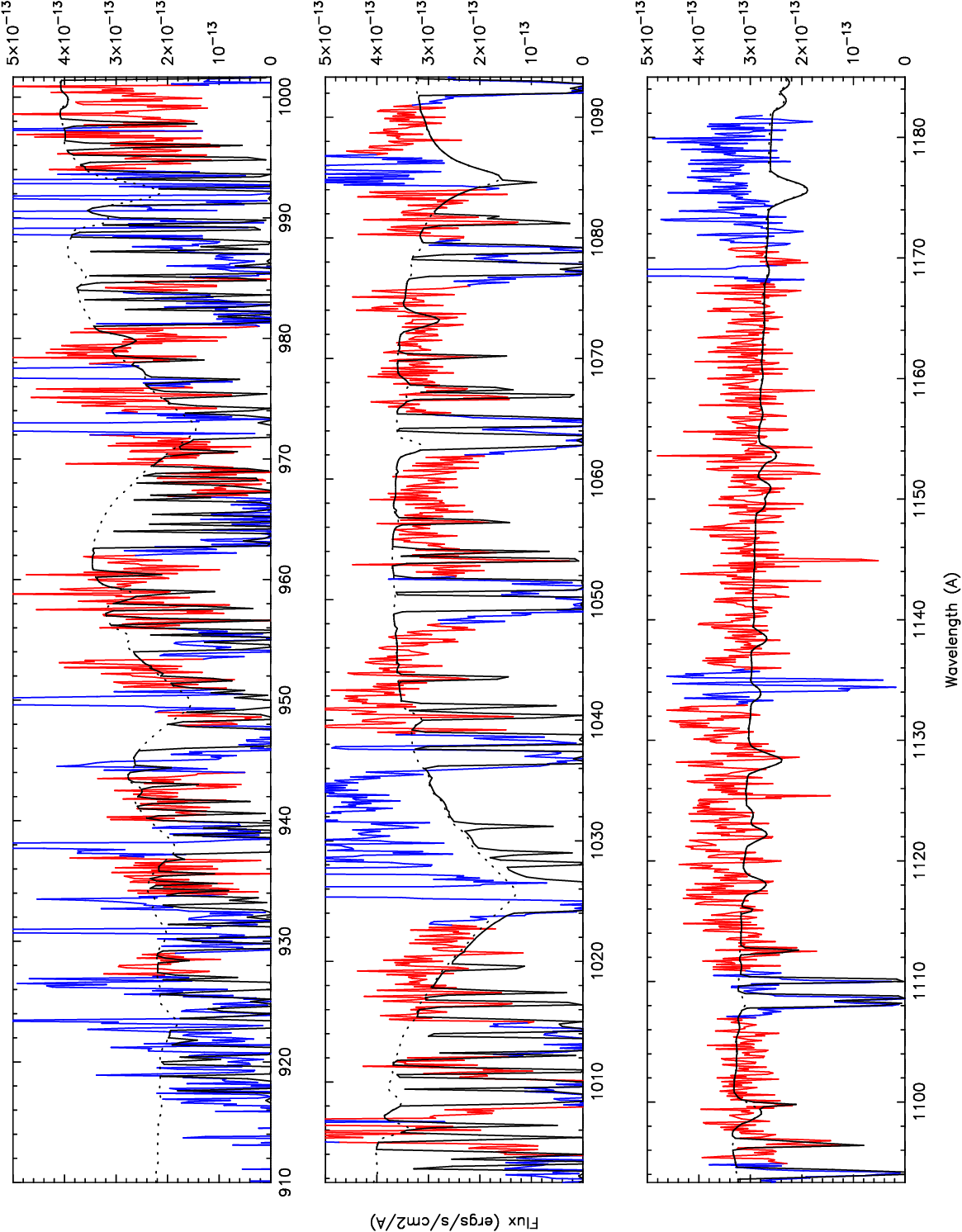}                          
\end{figure} 
\clearpage 

\begin{figure}
\plotone{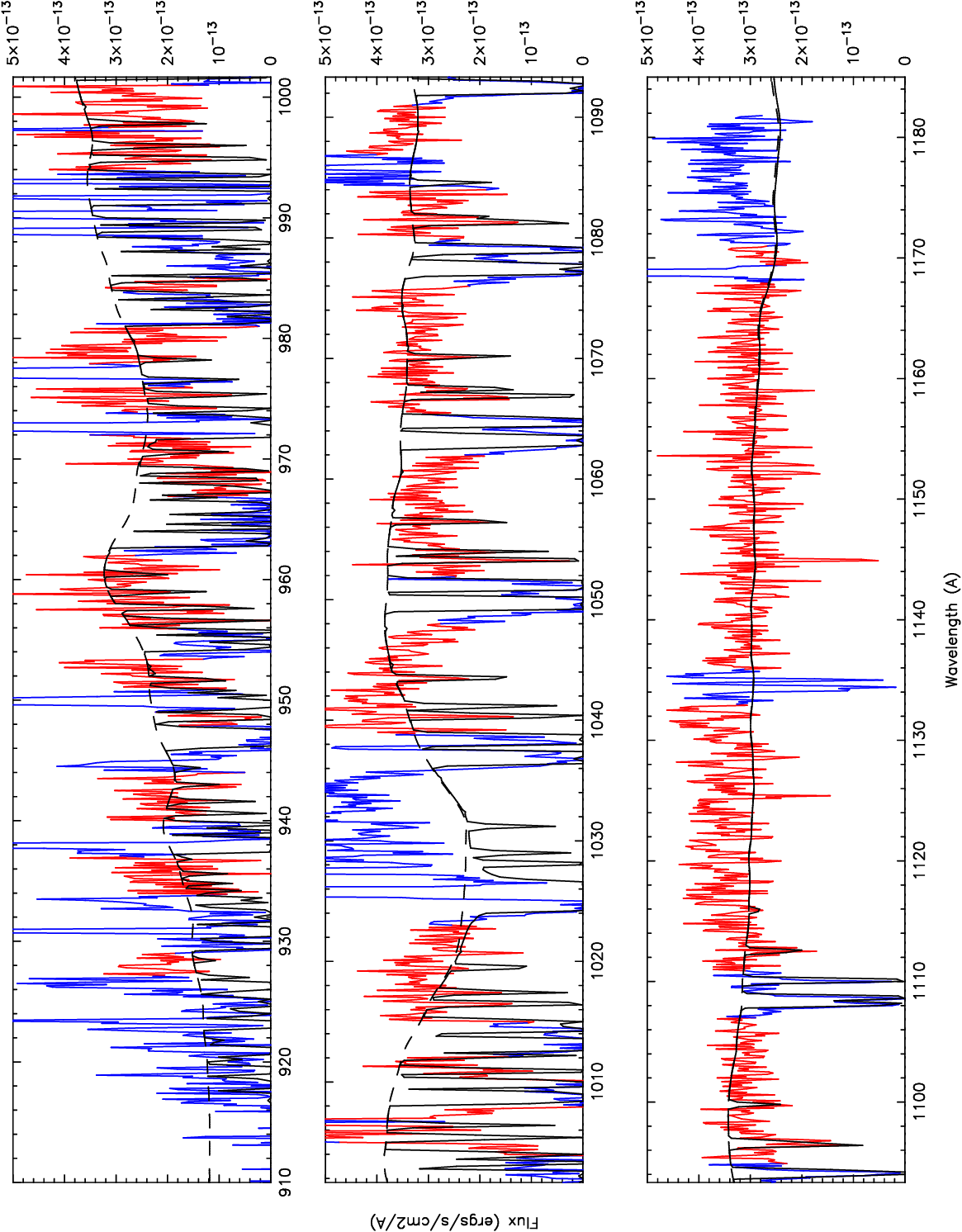}                          
\end{figure} 
\clearpage 

\begin{figure}
\plotone{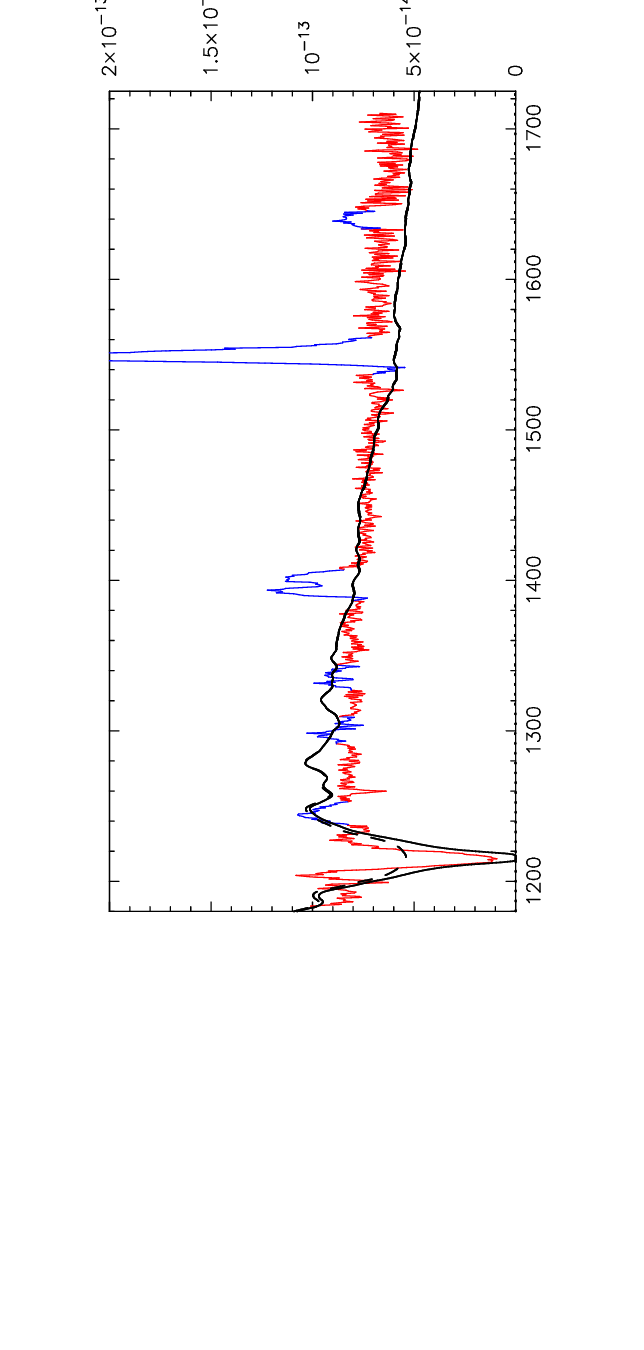}                          
\end{figure} 
\clearpage 

\begin{figure}
\plotone{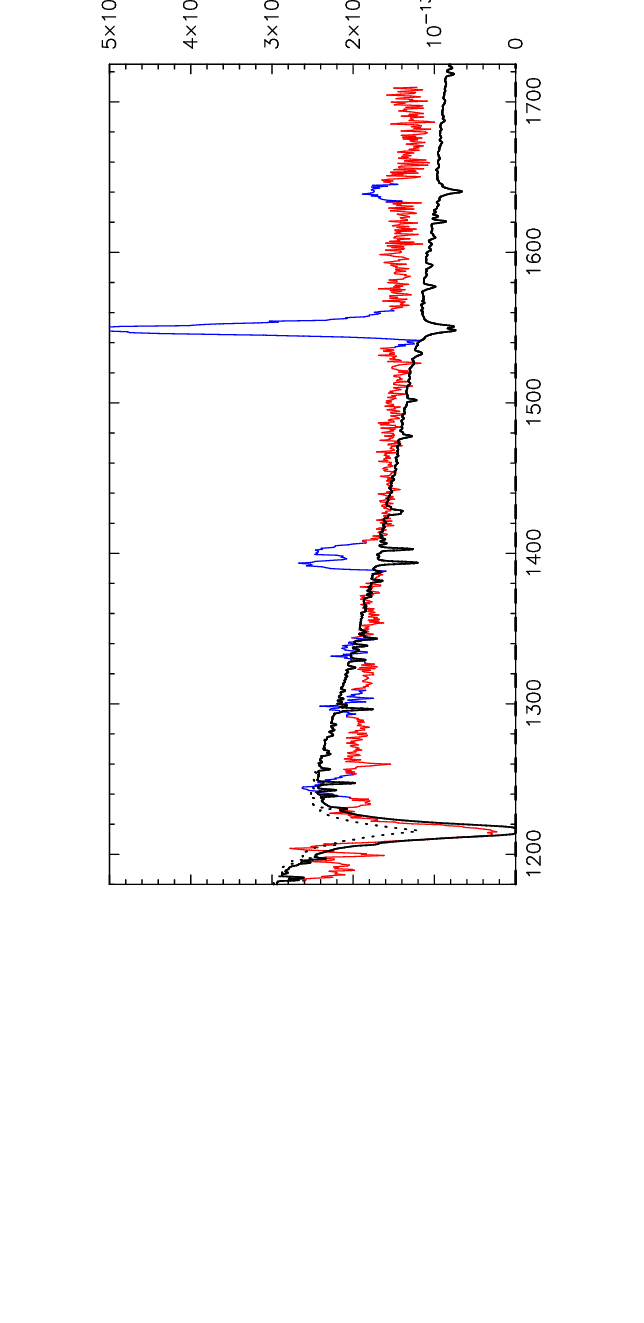}                          
\end{figure} 
\clearpage 

\begin{figure}
\plotone{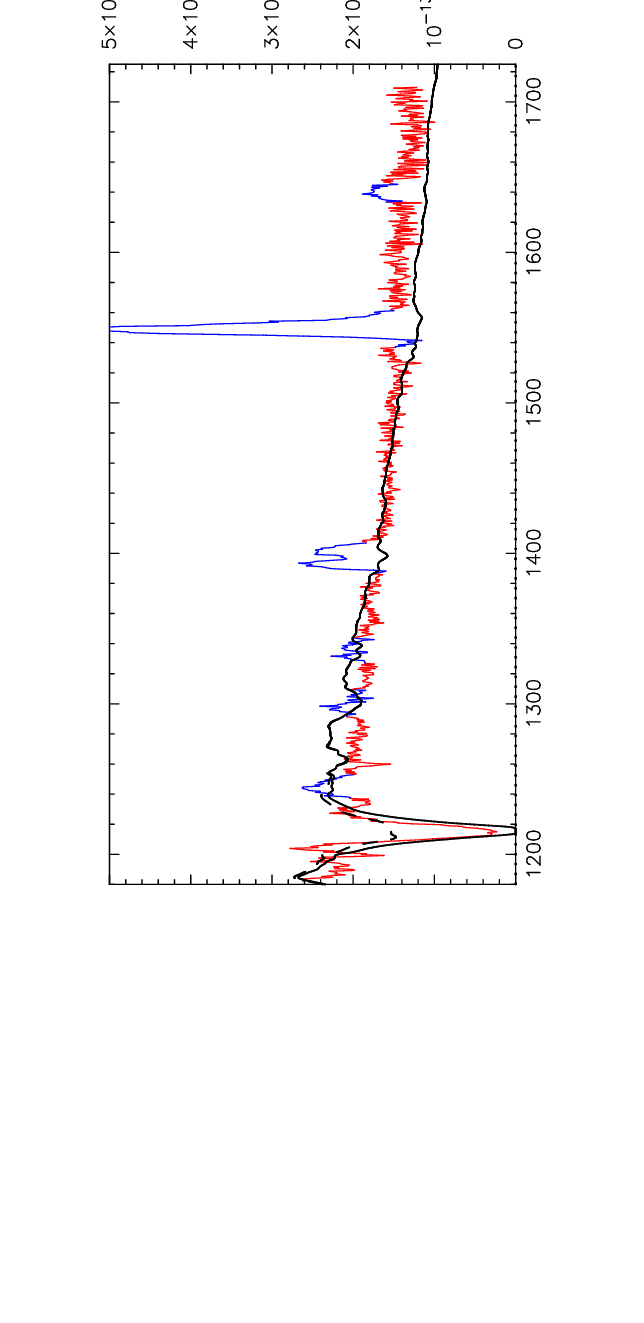}                          
\end{figure} 
\clearpage

\begin{figure}
\plotone{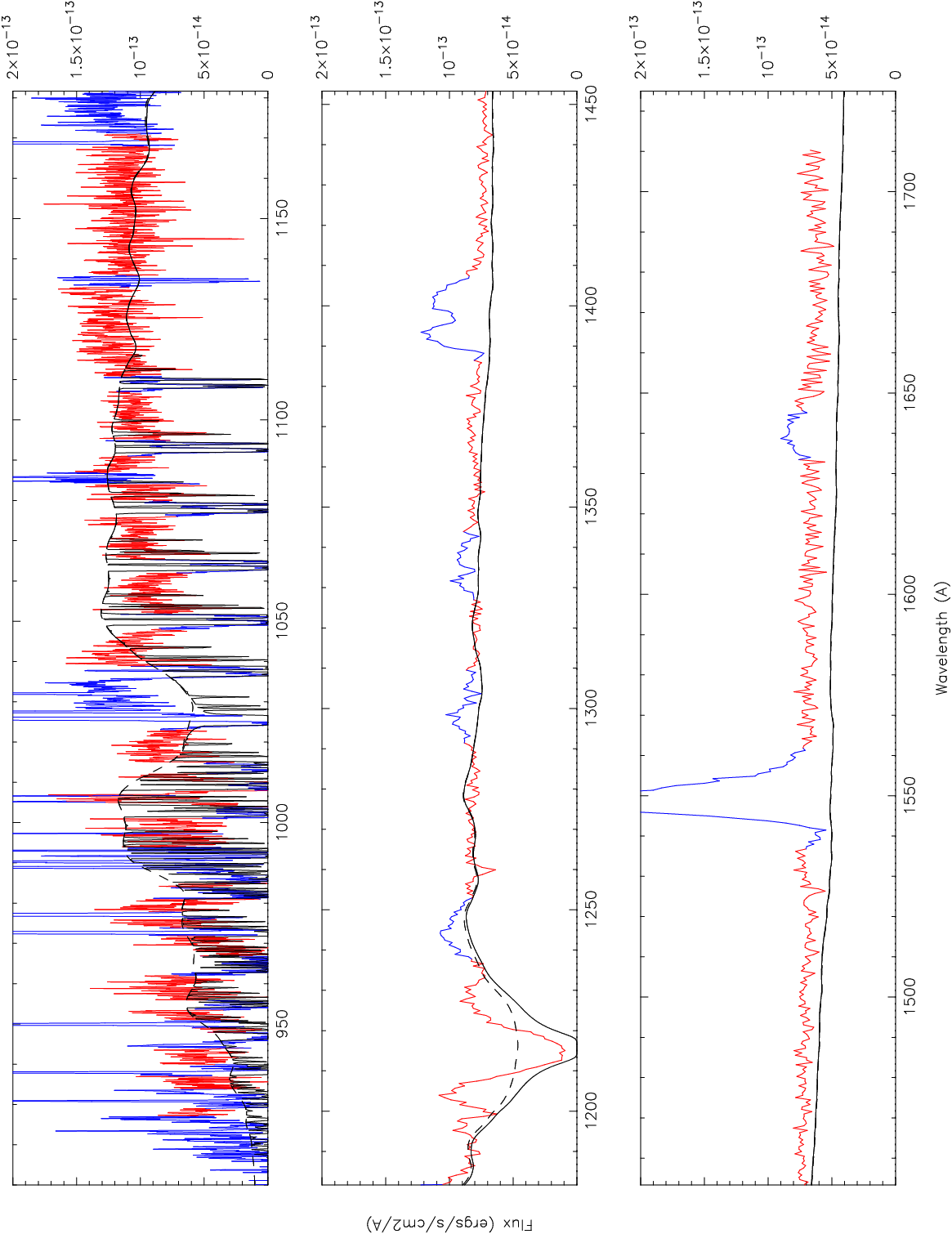}                          
\end{figure} 
\clearpage 

\begin{figure}
\plotone{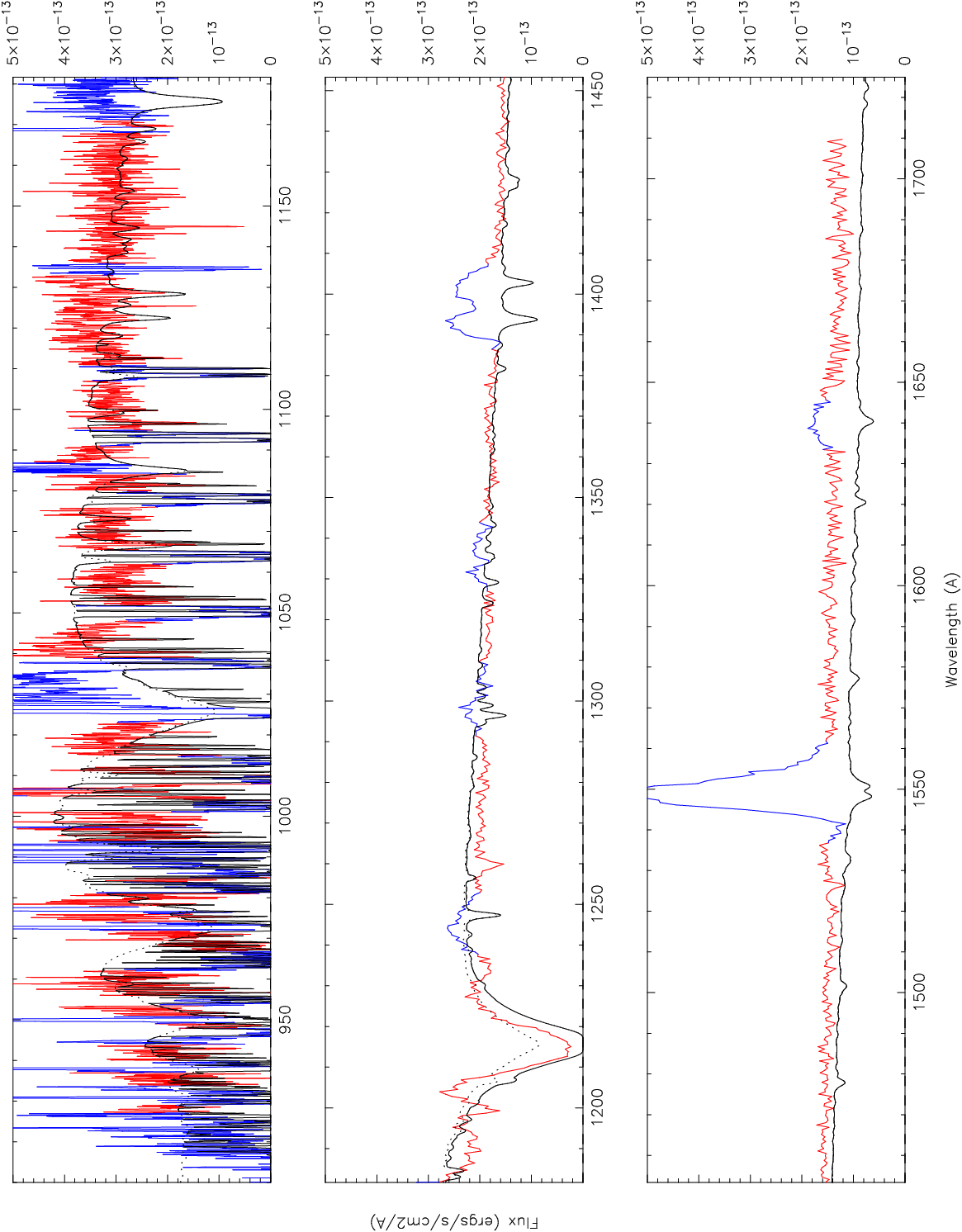}                          
\end{figure} 
\clearpage 

\begin{figure}
\plotone{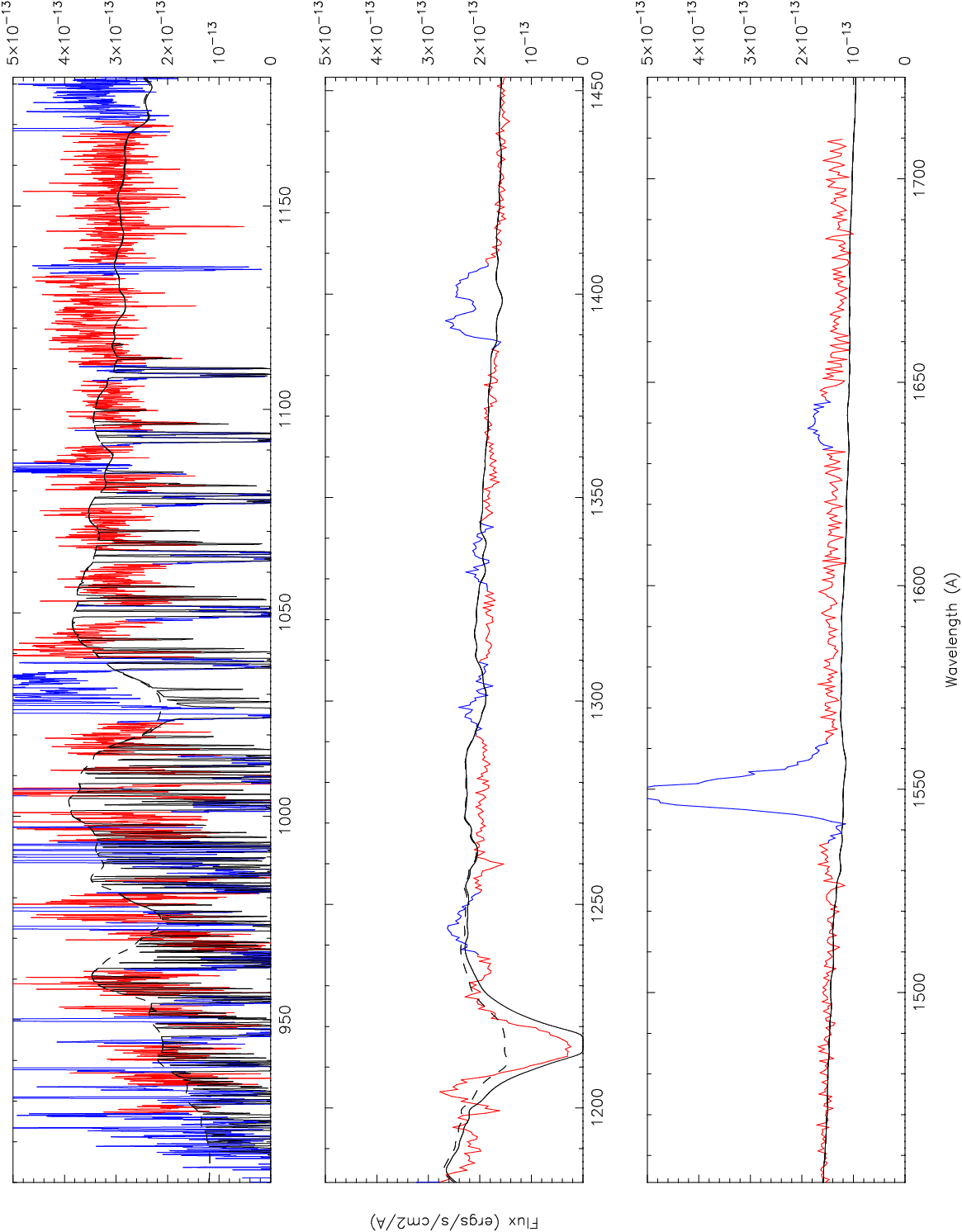}                          
\end{figure} 
\clearpage 

\begin{figure}
\plotone{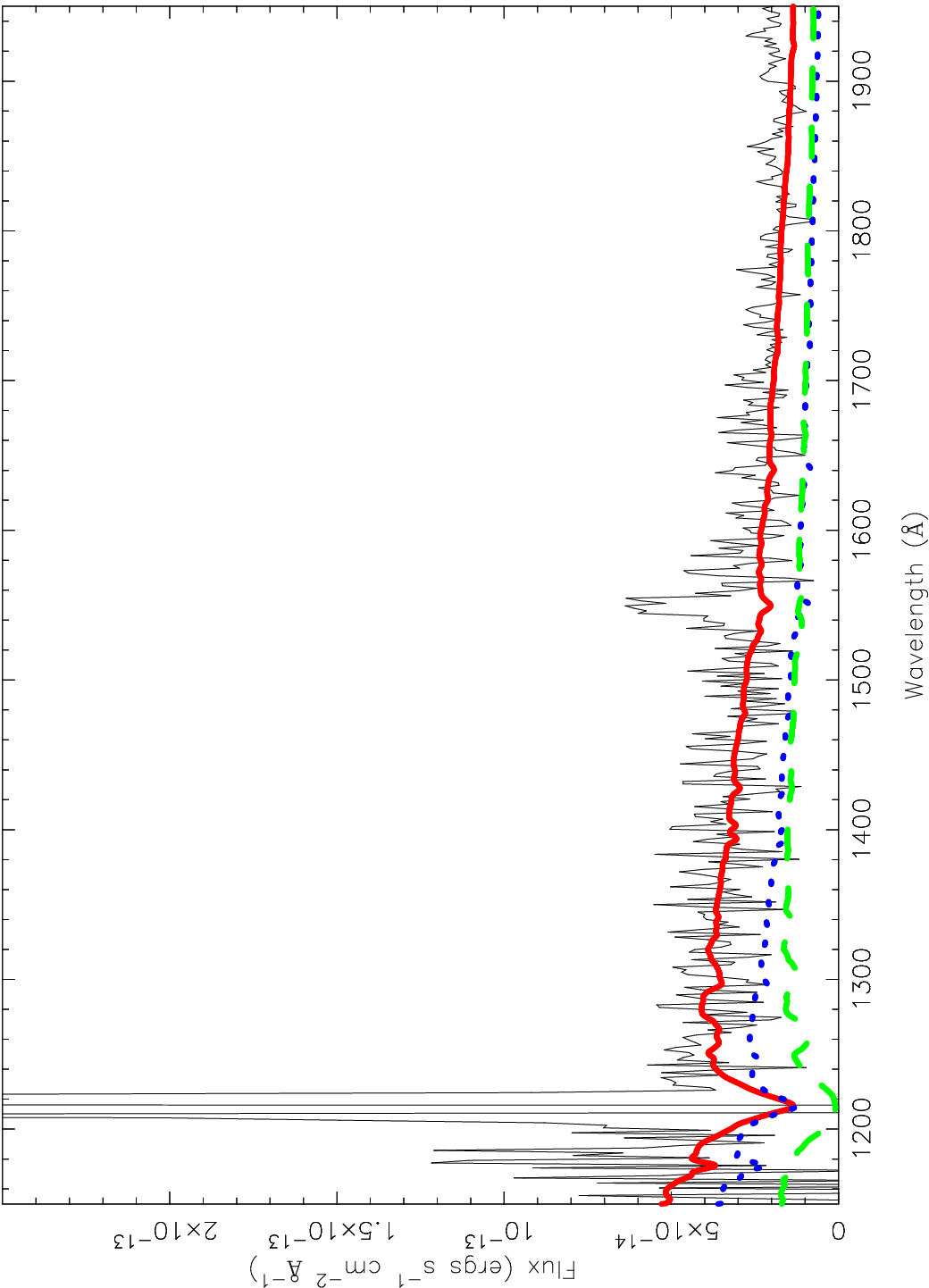}                          
\end{figure} 
\clearpage


\clearpage 

\begin{thebibliography}{}

\bibitem[Abgrall et al. (2000)]{abg00}
Abgrall, H., Roueff, E., \& Drira, I. 2000, A\&AS, 141, 297 

\bibitem[Bohlin et al. (1978)]{boh78}
Bohlin, R.C., Savage, B.D., \& Drake, J.F. 1978, \apj, 224, 132 

\bibitem[Bruch \& Engel (1994)]{bru94}
Bruch, A., \& Engel, A. 1994, AASS, 104, 79 

\bibitem[Cannizzo (1993)]{can93}
Cannizzo, J.K. 1993, \apj, 419, 780

\bibitem[Cannizzo (1998)]{can98}
Cannizzo, J.K. 1998, \apj, 493, 426


\bibitem[G\"ansicke et al. (1999)]{gan99}
G\"ansicke, B.T., Sion, E.M., Beuermann, K., Fabian, D.,
Cheng, F., Krautter, J. 1999, A\&A, 347, 178 

\bibitem[Godon et al. (2004)]{god04}
Godon, P., Sion, E.M., Cheng, F., Szkody, P., Long, K.S., \& Froning, C.S. 
2004, \apj, 612, 429   

\bibitem[Godon et al. (2006)]{god06}
Godon, P., Seward, L, Sion, E.M., \& Szkody, P. 2006, \aj, 131, 2634  

\bibitem[Hamada \& Salpeter (1961)]{ham61}
Hamada, T., \& Salpeter, E.E. 1961, \apj, 134, 683 

%
%

\bibitem[Honeycutt, Cannizzo, \& Robertson (1994)]{hon94} 
Honeycutt, R.K., Cannizzo, J.K., \& Robertson, J.W. 1994, \apj, 425, 835 

\bibitem[Honeycutt \& Robertson (1998)]{hon98}
Honeycutt, R.K., \& Robertson, J.W. 1998, \apj, 116, 1961 

\bibitem[Honeycutt \& Schlegel (1985)]{hon85}
Honeycutt, R.K., \& Schlegel, E.M. 1985, PASP, 97, 1189 

\bibitem[Hoard et al. (2003)]{hoa03}
Hoard, D.W., Szkody, P., Froning, C.S., Long, K.S., \&
Knigge, C. 2003, \apj, 126, 2473 

\bibitem[Hoard et al. (2004)]{hoa04}
Hoard, D.W., Linnell, A.P., Szkody, P., Fried, R.E., Sion, E.M.,
\& Wolfe, M.A. 2004, \apj, 604, 346  

\bibitem[Hubeny (1988)]{hub88}
Hubeny, I. 1988, Comput. Phys. Commun., 52, 103

\bibitem[Hubeny \& Lanz (1995)]{hub95}
Hubeny, I.,\& Lanz, T. 1995, \apj, 439, 875

\bibitem[Knigge et al. (2000)]{kni00}
Knigge, C., Long, K.S., Hoard, D.W., Szkody, P., \& Dhillon, 
V.S. 2000, \apj, 539, L49 

\bibitem[La Dous (1991)]{lad91}
La Dous, C. 1991, \aap, 252, 100

\bibitem[Massa \& Fitzpatrick (2000)]{mas00}
Massa, F., \& Fitzpatrick, E. 2000, \apjs, 126, 517

\bibitem[Mateo \& Szkody (1984)]{mat84}
Mateo, M., \& Szkody, P. 1984, \aj, 89, 863 

\bibitem[McCandliss (2003)]{mcc03}
McCandliss, S.R. 2003, \pasp, 115, 651 

\bibitem[Morton (2000)]{mor00} 
Morton, D.C. 2000, \apjs, 130, 403 

\bibitem[Morton (2003)]{mor03} 
Morton, D.C. 2003, \apjs, 149, 205 

\bibitem[Oke (1974)]{oke74}
Oke, J.B. 1974, \apjs, 27, 21 


\bibitem[Press et al. (1992)]{pre92}
Press, W.H., Teukolsky, S.A., Vetterling, W.T., Flannery, B.P.,
Numerical Recipes in Fortran 77, The Art of Scientific Computing,
Second Edition, 1992, Cambridge University Press 

\bibitem[Rana et al. (2005)]{ran05} 
Rana, V.R., Singh, K.P., Barrett, P.E., \& Buckley, D.A.H. 2005,
\apj, 625, 351 

%
\bibitem[Ritter \& Kolb (2003)]{rit03}
Ritter H., \& Kolb U. 2003, A\&A, 404, 301 
$<$http://physics.open.ac.uk/RKcat/RKcat\_AA.ps$>$

\bibitem[Sion et al. (2004)]{sio04}
Sion, E.M., Cheng, F., Godon, P., Urban, J., \& 
Szkody, P. 2004, \aj, 128, 1834 

\bibitem[Sion (1999)]{sio99}
Sion, E.M. 1999, \pasp, 111,532 

\bibitem[Sion et al. (2006)]{sio07}
Sion, E.M., Cheng, F., Godon, P., \& Szkody, P. 2007, \aj, submitted  


\bibitem[Szkody (1982)]{szk82}
Szkody, P. 1982, in Advances in Ultraviolet Astronomy: Four Years of 
{\it{IUE}} Research, NASA, 474

\bibitem[Szkody et al. (1981)]{szk81}
Szkody, P., Crosa, L., Bothun, G.D., Downes, R.A., \& Schommer, R.A. 1981,
\apj, 249, L61  

\bibitem[Szkody et al. (1988)]{szk88}  
Szkody, P., Downes, R., \& Mateo, M. 1988, PASP, 100, 362 


\bibitem[Verbunt (1987)]{ver87} 
Verbunt, F. 1987, A\&A, 71, 339 

\bibitem[Wade \& Hubeny (1998)]{wad98}
Wade, R.A., \& Hubeny, I. 1998, \apj, 509, 350

\bibitem[Warner (1982)]{war82}
Warner, B. 1982, Inf.Bull.Var.Stars, No.2175 

%
\bibitem[Warner (1995)]{war95}
Warner, B. 1995, Cataclysmic Variable Stars (Cambridge: Cambridge Univ.
Press)

%

\bibitem[Wood (1990)]{woo90}
Wood, M.A. 1990, PH.D. Thesis, University of Texas, Austin 

%
\end{thebibliography}
\end{document}